\documentclass[twoside]{article}
\usepackage{amsfonts,amssymb,amsbsy,textcomp,marvosym,picins,amsmath,caption,threeparttable,amsthm,subfigure}
\usepackage{eurosym,mathrsfs,fancyhdr,multicol,graphics,indentfirst,color,bm,upgreek,booktabs,graphicx}
\usepackage{tabularx}
\renewcommand{\thefootnote}{\fnsymbol{footnote}}
\def\footnoterule{\kern 1mm \hrule width 10cm \kern 2mm}

\allowdisplaybreaks
\catcode`@=11
\def\title#1{\vspace{3mm}\begin{flushleft}\vglue-.1cm\Large\bf\boldmath\protect\baselineskip=18pt plus.2pt minus.1pt #1
\end{flushleft}\vspace{1mm} }
\def\author#1{\begin{flushleft}\normalsize #1\end{flushleft}\vspace*{-4pt} \vspace{3mm}}
\def\address#1#2{\begin{flushleft}\vglue-.35cm${}^{#1}$\small\it #2\vglue-.35cm\end{flushleft}\vspace{-2mm}\par}

\catcode`@=11
\def\section{\@startsection{section}{1}{\z@}%
 {-3ex \@plus -.3ex \@minus -.2ex}%
 {2.2ex \@plus.2ex}%
{\normalfont\normalsize\protect\baselineskip=14.5pt plus.2pt minus.2pt\bfseries}}
\def\subsection{\@startsection{subsection}{2}{\z@}%
 {-3ex\@plus -.2ex \@minus -.2ex}%
 {2ex \@plus.2ex}%
{\normalfont\normalsize\protect\baselineskip=12.5pt plus.2pt minus.2pt\bfseries}}
\def\subsubsection{\@startsection{subsubsection}{3}{\z@}%
 {-2.2ex\@plus -.21ex \@minus -.2ex}%
 {1.4ex \@plus.2ex}
{\normalfont\normalsize\protect\baselineskip=12pt plus.2pt minus.2pt\sl}}

\begin{document}
\thispagestyle{empty}
\vspace*{-13mm}
\noindent {\small Zhang YW, Jin Z, Wang
ZJ {\it et al.} Summarization-Guided Assert Statement Generation.
JOURNAL OF COMPUTER SCIENCE AND TECHNOLOGY \ 33(1): \thepage--\pageref{last-page}
\ January 2018. DOI 10.1007/s11390-015-0000-0}
\vspace*{2mm}
\title{SAGA: Summarization-Guided Assert Statement Generation}

\author{Yu-Wei Zhang$^{1,2}$, Member, CCF, Zhi Jin$^{1,2,*}$, Fellow, CCF, IEEE, Ze-Jun Wang$^{1,2}$, Student Member, CCF, Ying Xing$^{3}$, Senior Member, CCF, and Ge Li$^{1,2,*}$, Senior Member, CCF, Member, IEEE, ACM}

\address{1}{Key Laboratory of High Confidence Software Technologies (Peking University), Ministry of Education, Beijing 100871, China}
\address{2}{School of Computer Science, Peking University, Beijing 100871, China}
\address{3}{School of Artificial Intelligence, Beijing University of Posts and Telecommunications, Beijing 100876, China}

\vspace{2mm}

\noindent E-mail: \{yuweizhang, zhijin, zejunwang\}@pku.edu.cn; xingying@bupt.edu.cn; lige@pku.edu.cn \\[-1mm]

\noindent Received July 15, 2018 [\textcolor{blue}{Month Day, Year}]; accepted October 14, 2018 [\textcolor{blue}{Month Day, Year}].\\[1mm]

\let\thefootnote\relax\footnotetext{{}\\[-4mm]\indent\ Regular Paper\\[.5mm]
\indent\ This work was supported by the National Natural Science Foundation of China under Grant Nos.~62072007, 62192733, 61832009, 62192731 and 62192730. \\[.5mm]
\indent\ $^*$Corresponding Author
\\[.5mm]\indent\ \copyright Institute of Computing Technology, Chinese Academy of Sciences 2021}

\noindent {\small\bf Abstract} \quad  {\small Generating meaningful assert statements is one of the key challenges in automated test case generation, which requires understanding the intended functionality of the tested code. Recently, deep learning-based models have shown promise in improving the performance of assert statement generation. However, existing models only rely on the test prefixes along with their corresponding focal methods, yet ignore the developer-written summarization. Based on our observations, the summarization contents usually express the intended program behavior or contain parameters that will appear directly in the assert statement. Such information will help existing models address their current inability to accurately predict assert statements. This paper presents a novel summarization-guided approach for automatically generating assert statements. To derive generic representations for natural language (i.e., summarization) and programming language (i.e., test prefixes and focal methods), we leverage a pre-trained language model as the reference architecture and fine-tune it on the task of assert statement generation. To the best of our knowledge, the proposed approach makes the first attempt to leverage the summarization of focal methods as the guidance for making the generated assert statements more accurate. We demonstrate the effectiveness of our approach on two real-world datasets when compared with state-of-the-art models.}

\vspace*{3mm}


\vspace*{4mm}

\baselineskip=15.8pt plus.2pt minus.2pt
\parskip=0pt plus.2pt minus0.2pt
\begin{multicols}{2}

\section{Introduction}

Software testing has been widely recognized as playing a critical role in improving software reliability during the software development life cycle (SDLC) \cite{1}. Effective unit testing is helpful to expose potential software faults early in SDLC to prevent the release of buggy software. However, writing high-quality unit test cases is a time-consuming and error-prone task in practice. To mitigate the manual costs of testing activities, extensive work has been devoted to the automatic generation of unit test cases \cite{2,3,4}. Even though these tools represent a notable achievement towards the goal of automated test case generation, several limitations have been highlighted by recent work in industrial settings \cite{5,6}. One major challenge lies in generating meaningful assert statements.

Recently, deep learning (DL) techniques have been applied to automated assert statement generation \cite{7,8}. Such DL-based models (e.g., ATLAS \cite{8}) take the test prefixes (i.e., test methods without any assert statements) along with corresponding focal methods (i.e., the basic units under test) as input. Specifically, these models are trained with a large corpus of paired test prefixes and focal methods, including method signatures and bodies. By learning semantic representations of the encoded input sequences, the trained models have the ability to automatically generate assert statements.

Nevertheless, generating meaningful assert statements is a tricky problem that requires a complete understanding of the intended functionality of the focal methods. The effectiveness of existing models is still limited due to a lack of useful contextual information (e.g., the summarization of focal methods). In an in-depth investigation of the developer-written summarization, we observe that some intent-related parameters within the assert statements can be directly discovered in the summarization contents but not in the source code implementations. Additionally, developer-written summarization conveys important information about the intended program behavior.

As shown in Fig.\ref{example1}, the statement (line 16) written by developer asserts that the return value of function \texttt{getTrueWindDirection} is equal to 234.5 within a positive delta 0.1. Specifically, the state-of-the-art model ATLAS only relies on the source code implementations (lines 5-15) to generate the recommended assert statement (line 17), which fails to predict the delta value. However, in the above example, the summarization of the focal method (lines 1-4) contains the specific delta value 0.1 (underlined at line 3) in its content. This indicates that the summarization may have explicitly given the intent-related parameter if such developer-written summarization content is available.

\begin{center}
\includegraphics[width=0.26\textwidth]{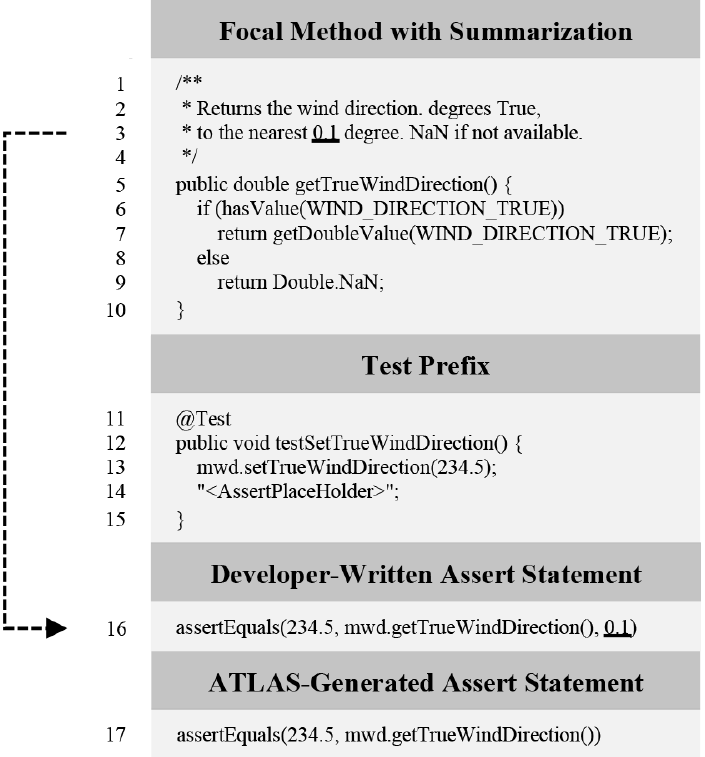}\\
\vspace{2mm}
\parbox[c]{8.3cm}{\footnotesize{Fig.1.~}  An example of developer-written assert statement containing parameter out of the code. }
\label{example1}
\end{center}

Considering another more complex real-world example shown in Fig.2, the ATLAS-generated result (line 30) correctly predicts the type of assert statement yet fails to capture the developer's intent from the content of the test prefix (lines 21-28) and focal method (lines 6-20). In this example, the developer asserts that the focal method \texttt{identifyOSXVersion} should return a more accurately identified version number (i.e., ``10.7.3'') of the operating system OS X at the resultant state according to the input string \texttt{userAgent}. Obviously, it is difficult for ATLAS to generate such a semantically correct assert statement without additional information pertaining to the intended functionality of the focal method. Likewise, the summarization written by the developer (lines 1-5) can be utilized to better understand the functionality of the focal method and capture the developer's intent.

\begin{center}
\includegraphics[width=0.36\textwidth]{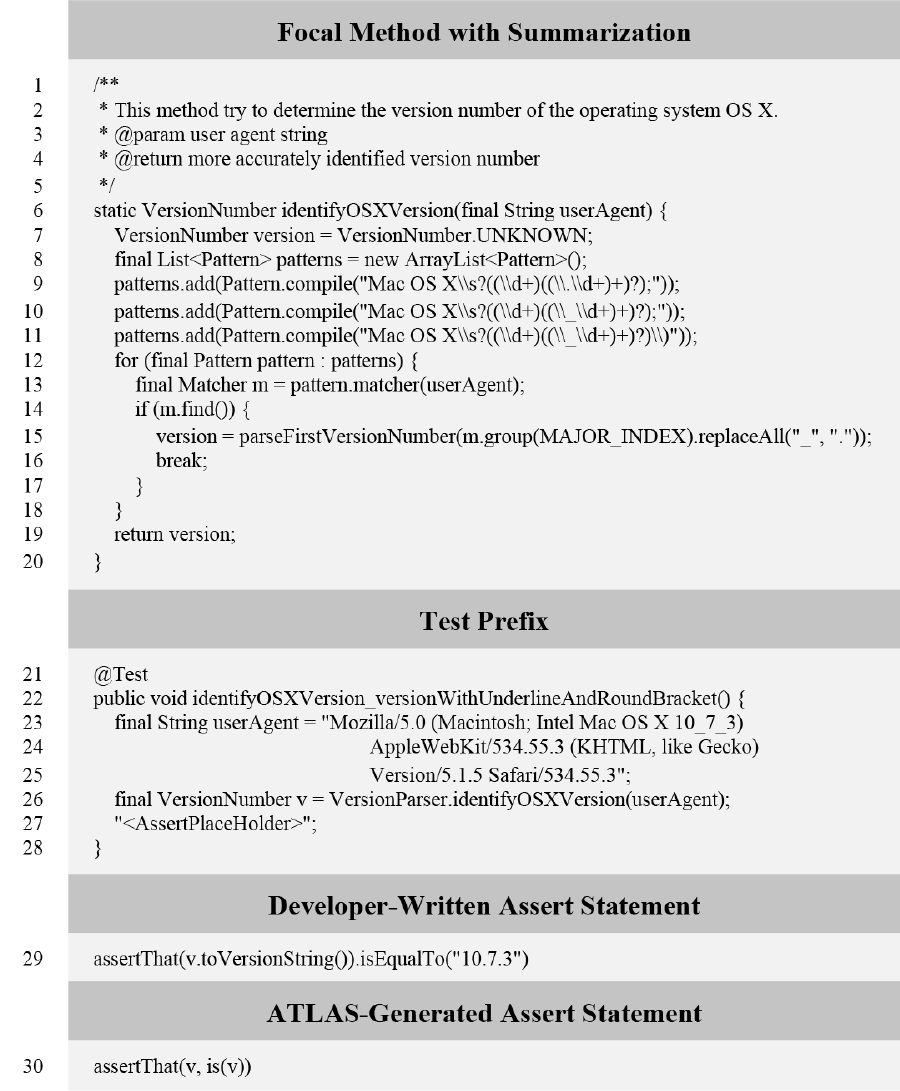}\\
\vspace{2mm}
\parbox[c]{8.3cm}{\footnotesize{Fig.2.~}  An example of ATLAS-generated assert statement failing to capture the developer's intent. }
\label{example2}
\end{center}

In summary, an effective approach for automated assert statement generation should not simply rely on the contents of source code to predict both the type and logical nature. Other contextual information (e.g., summarization) can also be utilized to assist the generation of correct assert statements. To that end, we present SAGA, a \textbf{S}umm\textbf{A}rization-\textbf{G}uided \textbf{A}ssert statement generation model, to address the limitations of existing neural generative approaches. The proposed model takes information from two modalities as input, which consist of the source code implementations (i.e., test prefix and focal method) written in programming language (PL) and the summarization contents written in natural language (NL). It needs to learn semantic representations of both PL and NL and correctly generate assert statements written in PL. Aiming to derive generic representations for NL and PL, we take advantage of the recently proposed model CodeT5 \cite{9}, a Text-To-Text Transfer Transformer (T5) \cite{10} architecture-based framework that leverages the NL-PL pairs to learn a better cross-modal alignment. We start with CodeT5 to train SAGA and then fine-tune it for the specialized downstream task (i.e., assert statement generation). In previous studies \cite{7,8}, the two code implementations are fed together into the model as a unified code snippet. Therefore, the model may have the burden of identifying the location of different code implementations. In contrast, the contents of test prefix, focal method, and summarization are isolated by special tokens in this paper and then fed to the model, which would provide SAGA with more information about different modalities for better learning the relationships between them. Our empirical investigations indicate that providing the summarization as guidance reinforces the performance of assert statement generation. By learning the semantic representations of NL and PL, SAGA can capture definitive information about relationships between the summarization contents and the source code implementations, thus aiding the generation of meaningful assert statements.

To evaluate the proposed approach, we adapt two real-world datasets \cite{8,11} to create our variant dataset named CAPS (\textbf{C}ode-\textbf{A}ssert \textbf{P}airs with \textbf{S}ummarization). Original datasets consist of paired source code (i.e., test cases mapped to corresponding focal methods) collected from large-scale open-source GitHub projects. To construct our adapted datasets, we discard the pairs for which we are not able to obtain the summarization of focal methods. Experimental results on the modified datasets demonstrate that SAGA is able to outperform the state-of-the-art approaches.

This paper makes the following contributions:
\begin{itemize}
  \item We present the first attempt at leveraging the developer-written summarization to guide the task of assert statement generation.
  \item We construct adapted datasets for assert statement generation that incorporate source code and summarization, which are publicly available in our online package$^{\footnotesize\textcircled{\tiny1}}$.
  \item We conduct an extensive evaluation on assert statement generation and demonstrate the effectiveness of using summarization for improving the model performance.
\end{itemize}
\let\thefootnote\relax\footnotetext{{}\indent\ $^{\footnotesize\textcircled{\tiny1}}$https://doi.org/10.5281/zenodo.7571911, Jan. 2023.}

The remainder of this paper is organized as follows. We describe the related work in Section~\ref{rel}. Section~\ref{met} introduces in detail the proposed approach. We outline the experimental setup in Section~\ref{exp}. Section~\ref{res} presents the results of our research. We disclose the threats to the validity of our approach in Section~\ref{thr}. Section~\ref{con} draws conclusions and indicates future directions.

\setcounter{figure}{2}
\begin{figure*}[b]
  \centering
  \includegraphics[width=\textwidth]{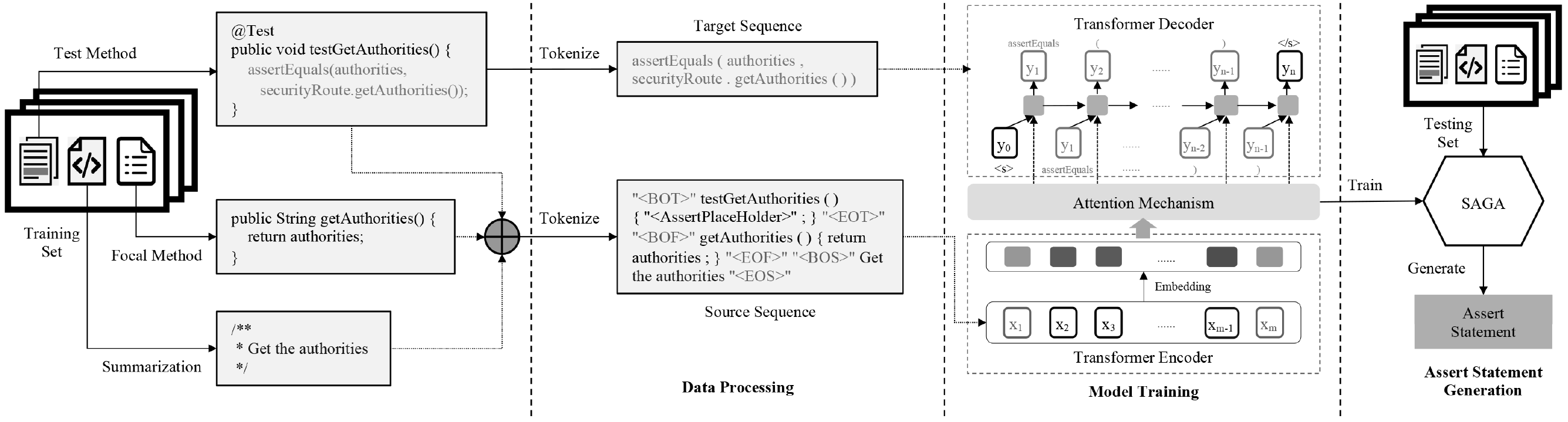}
  \caption{Overview framework of SAGA.}
  \label{fig3}
\end{figure*}

\section{Related Work}
\label{rel}

In recent years, there has been considerable interest in the automatic generation of assert statements. Numerous automated test generation tools have been proposed to synthesize assert statements using their own methods. EvoSuite \cite{4} applies a novel hybrid approach that generates and optimizes whole test suites towards satisfying a coverage criterion. It utilizes the system based on mutation and constraint solving to generate appropriate assert statements. Specifically, it introduces mutants into the system and attempts to generate assert statements that are capable of killing these mutants. Randoop \cite{2,3} is another automated tool that generates assert statements using feedback-directed random testing, a technique inspired by random testing that uses execution feedback gathered from executing test inputs as they are created to avoid generating redundant and illegal inputs. Essentially, a list of contracts, or pieces of logic that the code must follow, is used to guide the generation of assert statements. These contracts are very similar to developer-written assert statements. However, the contracts only provide the logic. Randoop creates a syntactically correct assert statement that tests the developer's provided logic pertaining to the test method. JQF \cite{12} combines fuzz testing and property-based testing to generate test cases, and the developer needs to manually write the test input when encountering an object as a test input.

With the advances of DL techniques, an increasing number of studies have so far utilized powerful DL models to tackle problems in the realm of software testing, such as bug localization \cite{13,14}, defect prediction \cite{15,16,17,18}, test case prioritization \cite{19,20,21,22}, and program repair \cite{23,24,25}. Such neural techniques have also shown promising results in automated assert statement generation. One such approach is ATLAS \cite{8}, which utilizes the recurrent neural network (RNN) to predict meaningful test oracles for given focal methods and test methods. Mastropaolo {\it et al.} \cite{26} investigated the performance of the T5 architecture on code-related tasks and found that the T5 model can be successfully applied to the assert statement generation task. Specifically, they first pre-trained a T5 model on a large corpus consisting of English sentences and source code, and then fine-tuned it on several downstream tasks including assert statement generation. Mastropaolo {\it et al.} \cite{27} further analyzed the benefits of pre-training and multi-task fine-tuning, and showed that the improved T5 model substantially boosted the performance on generating meaningful assert statements. Dinella  {\it et al.} \cite{28} proposed an end-to-end test generation approach TOGA that integrates neural test oracle generation with EvoSuite for bug detection, utilizing a Transformer-based model without relying on the unit's implementation. In contrast to DL-based approaches, Yu  {\it et al.} \cite{29} leveraged information retrieval (IR) techniques for generating assert statements, which is a two-stage approach including IR-based assert statement retrieval and adaptation. Furthermore, they introduced an integration strategy by combining the IR-based approach with ATLAS to improve its effectiveness.

Specifically, the aforementioned tools utilize handcrafted patterns or heuristics to infer assert statements for the test units. Instead, SAGA aims to mimic the behavior of developers when writing assert statements by using a DL-based approach. Furthermore, existing neural models simply rely on the source code implementations and lack the information of the developer-written summarization. Thus, SAGA makes the first attempt to leverage the summarization as complimentary information to accurately reflect the developer's intent to benefit the assert statement generation task.

\section{SAGA}
\label{met}

As shown in Fig.\ref{fig3}, the overall framework of SAGA mainly consists of three stages: data processing, model training, and assert statement generation. In this section, we first present an overview of the model architecture of SAGA and then detailed introduce each component of the proposed approach.

\subsection{Model Architecture}

In this paper, we adopt a sequence-to-sequence language model to learn semantic representations of both PL and NL for the task of assert statement generation. The model consists of an encoder that encodes the input sequences and a decoder that sequentially generates the expected assert statements, in which the encoder and decoder are both Transformers. Given an input token sequence $\bm{X}=(x_1, \dots ,x_m)$ (i.e., test prefix $+$ focal method $+$ summarization), SAGA first obtains the contextualized vector representations by projecting them into an embedded vector space through the embedding and positional encoding layer.
\begin{equation*}\label{eq1}
	\bm{X} = \text{Embedding}(\bm{X}) + \text{PositionalEncoding}(\bm{X}).
\end{equation*}

\setcounter{figure}{3}
\begin{figure*}[t]
  \centering
  \includegraphics[width=\textwidth]{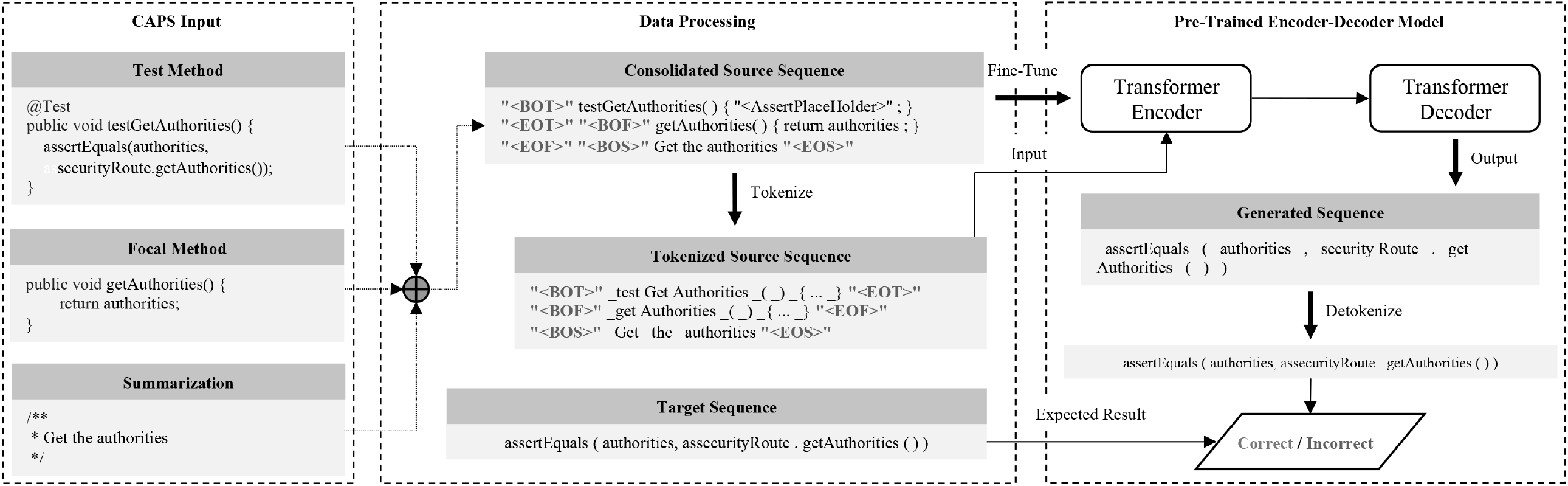}
  \caption{The pipeline of SAGA.}
  \label{fig4}
\end{figure*}

Then, the vector $\bm{X}$ is fed into the encoder to capture the long-term dependencies from different perspectives of the input sequences. The encoder comprises a stack of Transformer layers, each of which contains a multi-head self-attention layer followed by a position-wise fully connected feed-forward network. Instead of performing a single attention function, all attention blocks are split up into independent ``heads'' whose outputs are concatenated and linearly projected back onto a space with the initial dimensionality. Each individual attention block computes the scaled dot-product attention with different linear projections. The details are given by the following equations:
\begin{equation*}\label{eq2}
	\text{MultiHead}(\bm{Q},\bm{K},\bm{V}) = \text{Concat}(head_1,\dots,head_h)\bm{W}^O,
\end{equation*}
\begin{equation*}\label{eq3}
	head_i = \text{Softmax}\left(\frac{\bm{q}_i \bm{k}_i^{\rm{T}}}{\sqrt{d_k}}\right)\bm{v}_i, i=1, \ldots, h,
\end{equation*}
where $\bm{Q}$, $\bm{K}$, and $\bm{V}$ represent the matrices of query, key, and value, respectively, while $\bm{q}_i$, $\bm{k}_i$, and $\bm{v}_i$ represent their split matrices for $head_i$. Specifically, $\bm{W}^O$ denotes the weight matrix for linear transformation, and $d_k$ denotes the scaling factor for computing scaled dot-product attention.

The decoder in SAGA is a Transformer-based decoder that generates one new token at a time until a special stop token is reached. The decoder is similar in structure to the encoder except for the usage of the mask mechanism in multi-head attention, which forces to attend only to past tokens and avoids distraction and information leakage of the subsequent tokens in training. Followed by the masked multi-head attention layer is another multi-head attention that attends to both the past tokens and the feature representations learned by the encoder. Finally, the output of the final decoder layer is fed into a softmax layer to map target token scores into target token probabilities.

\subsection{Data Processing}
\label{dataset}

At this stage, SAGA prepares the collected source code and summarization in such a manner that it can be directly fed into the encoder-decoder model. SAGA first generates consolidated input for the CAPS dataset, which combines the three pieces of developer-written content into one sequence isolated by special tokens. The consolidation of the CAPS dataset consists of three major steps: 1) ignoring the comments, new line characters and redundant whitespaces within the body of test and focal methods, and removing the comment-related symbols (e.g., ``$/*$'') within the summarization; 2) appending the summarization, signature and full body of the focal method to the end of the test method; 3) replacing the entire assert statement from the test method with the unique token ``\texttt{<AssertPlaceHolder>}''. The \texttt{Consolidated Source Sequence} in Fig.\ref{fig4} shows an example of CAPS input where tokens are separated by single whitespace. The input sequence involves the test prefix, focal method, and summarization, which are isolated by different abbreviate tokens (e.g., ``\texttt{<BOS>}'' denotes the beginning of summarization). And the \texttt{Target Sequence} in Fig.\ref{fig4} is a single assert statement that the decoder is expected to generate. Then, SAGA uses the sentence-piece tokenizer \cite{30} to divide every word token into a sequence of sub-word tokens for alleviating the open-vocabulary problem \cite{31}. In this manner, the tokens with low frequency can be synthesized more easily, thus making the assert statement generation task more tractable.

\subsection{Model Training}

As illustrated in Fig.\ref{fig4}, the training pipeline of SAGA takes the generated CAPS dataset as input and works on three different pieces of developer-written content: 1) the test method that contains information on how to test the focal method; 2) the full context of the focal method; 3) the summarization of the focal method written in NL. We begin with the pre-trained model serving as the reference architecture for the proposed SAGA framework. Firstly, the previous data processing step tokenizes the input CAPS dataset. We then perform the fine-tuning on the task of assert statement generation. At the final step, the encoder in SAGA encodes the tokenized source sequence, and the decoder sequentially predicts the assert statement.

\subsubsection{Pre-Training}

Based on the empirical findings from existing datasets, we believe that developer-written summarization can be beneficial to models specialized in the assert statement generation task. Moreover, compared to the other two code implementations, we consider the summarization as a different modality. In order to learn generic representations for NL and PL, we leverage the state-of-the-art model CodeT5 as the starting point to train SAGA. That is, we can take the learned parameters of a pre-trained model and use them as initialization for SAGA. The goal of SAGA is to automatically synthesize an assert statement for the given test prefix, focal method, and summarization. We formulate this task as a text-to-text prediction, which is consistent with CodeT5's design. Thus, by using the learned parameters pre-trained on the colossal clean crawled corpus, SAGA is able to learn many generic patterns that can be directly applied to the task of generating assert statements. In addition, we isolate different input modalities with special tokens (as described in Section~\ref{dataset}), which would further benefit SAGA in using CodeT5 to learn the relationship between NL and PL.

\subsubsection{Fine-Tuning}

At this stage, we fine-tune the pre-trained model for the task of generating assert statements. The fine-tuning techniques can optimize the pre-trained parameters to make them more suitable for the downstream tasks. Specifically, we represent the assert statement generation task in a ``text-to-text'' format, where the input is a consolidated sequence of test prefix, focal method, and summarization, and the output is the expected assert statement. The fine-tuning process is performed using the training corpus of CAPS dataset $D$, and each instance within $D$ can be formally represented as a pair $D_i = \{c, a\}$, where $c=(t, f, s)$ comprises the test prefix $t$, the corresponding focal method $f$ and the summarization $s$, and $a$ denotes the developer-written assert statement. The fine-tuning objective is to minimize the cross-entropy loss by learning the mapping $c \rightarrow a$ as a conditional probability $p(a|c)$.

\subsection{Assert Statement Generation}

To sum up, the encoder learns representations of every sub-word token in the input source sequence using all input instances in the training corpus, essentially encoding the whole input information in every input sub-word token representation. The self-attention mechanism allows the decoder to attend to all previously generated sub-word tokens and decide on generating the correct token at the correct place. During inference, SAGA uses beam search to generate the assert statement sequentially. Once the decoder reaches the stop token, SAGA outputs the top-ranked sequence in the beam search. SAGA then detokenizes the sequence of sub-word tokens to restore the original sequence. Finally, we compare the detokenized sequence with the target sequence to determine whether the SAGA-generated assert statement exactly matches the developer-written one.

\section{Experimental Setup}
\label{exp}

\subsection{Experimental Subjects}
\label{subjects}

During the dataset construction process, our goal is to map the test methods to their corresponding focal methods. For this task, we first mine a 100K sample of public GitHub repositories written in the Java programming language, which has been used in the previous studies \cite{8,11}. Next, we parse the selected projects and extract all the declared methods with their associated metadata (e.g., annotations, signatures, and variables) using Spoon \cite{32}. The parsed code will be utilized for identifying focal methods as well as augmenting the focal methods with summarization. Finally, for a particular test method, we map it to the corresponding focal method for deriving the CAPS dataset.

\subsubsection{Test and Focal Method Mapping}

In this stage, we establish the test-to-code traceability links (i.e., mapping the corresponding focal method to each test method) for the extracted methods. To this aim, we introduce the following hybrid heuristic strategy in this paper:
\begin{itemize}
	\item \textbf{Naming Convention (NC):} Considering the intention behind NC, test method names are often similar to the corresponding focal methods. Therefore, the first heuristic strategy attempts to match the test method with a focal method having a name that matches, after removing the possible \texttt{"Test"} prefix or suffix. If the names match exactly, the focal method is correctly identified for the test method.
	\item \textbf{Static Call Graph (SCG):} The NC technique would fail to identify the focal method when no test method name contains its name. To address the drawback of NC, we then use SCG to aid the mapping process if the previous heuristic strategy does not identify any focal method. The second heuristic strategy hypothesizes that we can derive the focal method by inspecting method invocations in the test methods. To identify the focal method of a particular test method, we begin by collecting all production classes that are the destination of an outgoing method invocation within the test method and selecting the most referenced production class as the focal class. Then, we compute the intersection between the list of method invocations within the test method and the list of methods declared within the focal class by querying the complete signature string. If the intersection is a unique method, then we select the method as the focal method.
\end{itemize}

\setcounter{table}{0}
\tabcolsep 9pt
\renewcommand\arraystretch{1.3}
\begin{table*}[!htb]
\centering
\caption{\label{capss} Detailed Statistics of the Two Adapted CAPS Datasets}\vspace{-2mm}
{\footnotesize
\resizebox{\textwidth}{!}{
\begin{tabular}{ccccccccccccc}
\hline\hline\hline
\textbf{Dataset} & \multicolumn{3}{c}{\textbf{Split} (\# of instances)} & \multicolumn{3}{c}{\textbf{Source Code Length} (\# of tokens)} & \multicolumn{3}{c}{\textbf{Summarization Length} (\# of tokens)} & \multicolumn{3}{c}{\textbf{Assert Statement Length} (\# of tokens)} \\
    \cmidrule(rl){2-4}
    \cmidrule(rl){5-7}
    \cmidrule(rl){8-10}
    \cmidrule(rl){11-13}
& Training & Validation & Testing & MaxL & MinL & AvgL & MaxL & MinL & AvgL & MaxL & MinL & AvgL \\
  \hline
  $\mathbf{CAPS}_A$ & 62386 & 7756 & 7789 & 984 & 11 & 131.8 & 659 & 3 & 31.0 & 747 & 4 & 12.8  \\
  $\mathbf{CAPS}_M$ & 93246 & 11492 & 12971 & 982 & 16 & 182.5 & 726 & 3 & 36.2 & 529 & 3 & 13.1 \\
  \hline\hline\hline
  \end{tabular}}
}
\end{table*}
\baselineskip=18pt plus.2pt minus.2pt
\parskip=0pt plus.2pt minus0.2pt

\subsubsection{CAPS Dataset Construction}

After identifying the corresponding focal methods for all the test methods, we further filter the focal methods without developer-written summarization to construct the CAPS dataset. Specifically, CAPS is a corpus of test prefixes, corresponding focal methods with summarization, and assert statements. Additionally, due to the possibility of cloning methods across different GitHub repositories, we further exclude duplicated instances to prevent the same instance appearing in both the training and testing sets. After preprocessing, we create two adapted CAPS datasets (referred to as $\mathbf{CAPS}_A$ and $\mathbf{CAPS}_M$) by modifying ATLAS \cite{8} and Method2Test \cite{11}, respectively. As mentioned in the previous study \cite{8}, the original ATLAS dataset is constructed in a simplified way that excludes some challenging cases (i.e., the assert statements that contain tokens absent from the input contents) for generation. In this paper, $\mathbf{CAPS}_A$ and $\mathbf{CAPS}_M$ include the cases of assert statements with unknown tokens. $\mathbf{CAPS}_A$ and $\mathbf{CAPS}_M$ contain a total of 77931 and 117709 unique instances, respectively. Next, we further split each dataset into training, validation, and testing sets by the ratio of 8:1:1. The dataset split is performed carefully by taking into account possible data leakage. To be specific, any two instances belonging to the same GitHub repository cannot be put in two different sets (e.g., one in training and the other in testing). In other words, all the instances belonging to the same GitHub repository will be put in the same set. Table~\ref{capss} reports the detailed statistics of the two modified datasets, where MaxL, MinL, and AvgL denote the maximum length, the minimum length, and the average length, respectively.

\subsection{Experimental Design}
\label{design}

We conduct experiments on the two adapted CAPS datasets to evaluate the effectiveness of SAGA. In this paper, we compare SAGA with the following five baselines that are related to our work:
\begin{itemize}
  \item \textbf{TestNMT} \cite{7}: an experimental approach to test generation using an RNN-based NMT model, allowing a developer to generate an approximate test for a given function.
  \item \textbf{ATLAS} \cite{8}: a DL-based model that uses the RNN encoder-decoder with copy-attention mechanism to generate assert statements.
  \item \textbf{T5} \cite{26}: a pre-trained model that is fine-tuned using the ATLAS dataset \cite{8} for the assert statement generation task.
  \item \textbf{T5-Extension} \cite{27}: an extended version of the T5 model \cite{26} paying particular attention at the role played by pre-training and multi-task fine-tuning on the model's performance.
  \item \textbf{integration} \cite{29}: an IR-based approach combined with ATLAS to enable more powerful assert statement generation.
\end{itemize}

\let\thefootnote\relax\footnotetext{{}\indent\ $^{\footnotesize\textcircled{\tiny2}}$https://huggingface.co/Salesforce/codet5-small, May. 2022.}

We initialize SAGA with the pre-trained CodeT5-small checkpoint$^{\footnotesize\textcircled{\tiny2}}$ from the Huggingface's website. We adopt the same architecture as T5 \cite{10} model, consisting of 8-headed attention and six layers in both the encoder and decoder. We set the maximum source and target sequence lengths both to 512 and the batch size to 256. For the implementation of baselines, we reimplement TestNMT and ATLAS with the same architectures and hyper-parameters described in the relevant papers using \texttt{OpenNMT-py} \cite{33}. As for T5 and T5-Extension, we use the publicly released checkpoints. As for integration, we download the available source code provided by the authors. To make a fair comparison, we uniformly use the training set of the CAPS dataset to train or fine-tune baselines and SAGA on the task of assert statement generation respectively. During the training or fine-tuning step, we train each corresponding model for a maximum of 100 epochs. After each epoch, we compute the loss on the validation set and save the model with the minimum validation loss. To avoid the over-fitting issue, we perform early stopping if the validation performance does not improve for five consecutive epochs. During testing, we use a beam search and set the beam size to five. Finally, we evaluate the trained model on the testing set and report the comparison results in this paper. We conduct experiments on four Nvidia GTX 1080Ti GPUs.

\subsection{Experimental Metrics}
\label{metric}

To quantitatively compare the performance of SAGA with the baselines, we choose the following three widely used metrics \cite{7,8,29}.
\begin{itemize}
  \item \textbf{Accuracy. } This paper uses the top-1 accuracy to measure the performance of the proposed approaches. When the generated assert statement matches exactly with the developer-written assert statement, it is correct. Otherwise, it is incorrect.
  \item \textbf{BLEU. } The BLEU (Bilingual Evaluation Understudy) \cite{34} score is a variant of the precision metric widely used to assess the quality of NMT systems. This metric can calculate the similarity by computing the n-gram precision of a candidate sentence to the reference sentence, with a penalty for the overly short length. In this paper, we report the BLEU-4 score.
  \item \textbf{ROUGE. } ROUGE (Recall-Oriented Understudy for Gisting Evaluation) \cite{35} formally calculates an n-gram recall between a candidate sentence and a set of reference sentences. In this paper, we present the value of ROUGE-L, which computes the F-measure based on the longest common subsequence.
\end{itemize}

\setcounter{table}{1}
\tabcolsep 9pt
\renewcommand\arraystretch{1.3}
\begin{table*}[!htb]
\centering
\caption{\label{rq1a} Comparison Results of the Three Metrics for RQ1}\vspace{-2mm}
{\footnotesize
\begin{tabular}{lrrrrrr}
\hline\hline\hline
\textbf{Model} & \multicolumn{3}{c}{$\mathbf{CAPS}_A$} & \multicolumn{3}{c}{$\mathbf{CAPS}_M$} \\
    \cmidrule(rl){2-4}
    \cmidrule(rl){5-7}
& \multicolumn{1}{c}{\textbf{Accuracy}} & \multicolumn{1}{c}{\textbf{BLEU-4}} & \multicolumn{1}{c}{\textbf{ROUGE-L}} & \multicolumn{1}{c}{\textbf{Accuracy}} & \multicolumn{1}{c}{\textbf{BLEU-4}} & \multicolumn{1}{c}{\textbf{ROUGE-L}}  \\
  \hline
  \textbf{TestNMT} & 9.5\% & 21.74 & 60.95 & 1.1\% & 2.87 & 46.04 \\
    \textbf{ATLAS} & 18.0\% & 28.70 & 70.01 & 7.6\% & 14.22 & 60.36 \\
    \textbf{T5} & 9.1\% & 26.44 & 43.94 & 2.1\% & 20.31 & 49.74 \\
    \textbf{T5-Extension} & 23.8\% & 33.02 & 72.15 & 7.6\% & 21.86 & 57.85 \\
    \textbf{integration} & 37.2\% & 59.73 & 79.51 & 14.1\% & 22.78 & 62.26 \\
    \textbf{SAGA} & \textbf{53.1\%} & \textbf{75.56} & \textbf{85.96} & \textbf{19.8\%} & \textbf{39.17} & \textbf{65.15} \\
  \hline\hline\hline
  \end{tabular}
}
\end{table*}
\baselineskip=18pt plus.2pt minus.2pt
\parskip=0pt plus.2pt minus0.2pt

\section{Results and Analysis}
\label{res}

In this section, we present the experimental results for measuring the performance of SAGA and answering the following three research questions (RQs):
\begin{itemize}
  \item \textbf{RQ1}: How does SAGA perform compared with the state-of-the-art baselines?
  \item \textbf{RQ2}: What is the effectiveness of the developer-written summarization on the task of assert statement generation?
  \item \textbf{RQ3}: What is the quality of the generated incorrect assert statements?
\end{itemize}

\subsection{Answering RQ1}

To answer this question, we compare SAGA with five baselines on two adapted datasets. We remove the summarization contents from the CAPS corpus when training the baselines. In particular, we observe that each class name within the fine-tuning datasets of T5 and T5-Extension is preceded by its complete package name (e.g., the \textbf{String} class is tokenized as \textbf{java . lang . String}). Thus, we also apply such a modification to the two datasets when training the two baselines.

\setcounter{table}{2}
\tabcolsep 9pt
\renewcommand\arraystretch{1.3}
\begin{table*}[b]
\centering
\caption{\label{rq1b} Detailed Statistics of Each Assert Type}\vspace{-2mm}
{\footnotesize
  \resizebox{\textwidth}{!}{
  \begin{tabular}{clrrrrrrrrr}
    \hline\hline\hline
    \textbf{Dataset} & \multicolumn{1}{c}{\textbf{Model}} & \multicolumn{1}{c}{\textbf{True}} & \multicolumn{1}{c}{\textbf{False}} & \multicolumn{1}{c}{\textbf{Null}} & \multicolumn{1}{c}{\textbf{NotNull}} & \multicolumn{1}{c}{\textbf{Equals}} & \multicolumn{1}{c}{\textbf{Same}} & \multicolumn{1}{c}{\textbf{ArrayEquals}} & \multicolumn{1}{c}{\textbf{That}} & \multicolumn{1}{c}{\textbf{Other}} \\
    \hline
    $\mathbf{CAPS}_A$ & \textbf{TestNMT} & 148 (12.8\%) & 12 (2.9\%) & 64 (17.4\%) & 83 (21.7\%) & 377 (9.8\%) & 0 (0\%) & 18 (12.5\%) & 38 (2.7\%) & - \\
    & \textbf{ATLAS} & 207 (17.9\%) & 55 (13.1\%) & 105 (28.6\%) & 198 (51.8\%) & 687 (17.9\%) & 4 (3.8\%) & 36 (25.0\%) & 107 (7.7\%) & - \\
    & \textbf{T5} & 146 (12.6\%) & 15 (3.6\%) & 58 (15.8\%) & 75 (19.6\%) & 329 (8.6\%) & 0 (0\%) & 9 (6.3\%) & 80 (5.8\%) & - \\
    & \textbf{T5-Extension} & 342 (29.6\%) & 76 (18.1\%) & 112 (30.5\%) & 146 (38.2\%) & 903 (23.6\%) & 9 (8.6\%) & 37 (25.7\%) & 227 (16.3\%) & - \\
    & \textbf{integration} & 469 (40.6\%) & 116 (27.7\%) & 163 (44.4\%) & 221 (57.9\%) & 1404 (36.7\%) & 38 (36.2\%) & 60 (41.7\%) & 424 (30.5\%) & - \\
    & \textbf{SAGA} & 663 (57.4\%) & 221 (52.7\%) & 218 (59.4\%) & 266 (69.6\%) & 1935 (50.5\%) & 54 (51.4\%) & 72 (50.0\%) & 704 (50.7\%) & - \\
    $\mathbf{CAPS}_M$ & \textbf{TestNMT} & 0 (0\%) & 1 (0.2\%) & 0 (0\%) & 1 (0.2\%) & 136 (2.4\%) & 0 (0\%) & 0 (0\%) & 0 (0\%) & 11 (1.6\%) \\
    & \textbf{ATLAS} & 77 (5.0\%) & 47 (8.8\%) & 123 (24.9\%) & 181 (28.4\%) & 349 (6.2\%) & 0 (0\%) & 24 (8.6\%) & 63 (2.1\%) & 120 (17.8\%) \\
    & \textbf{T5} & 45 (2.9\%) & 5 (1.0\%) & 16 (3.2\%) & 3 (0.5\%) & 130 (2.3\%) & 0 (0\%) & 10 (3.6\%) & 22 (0.7\%) & 36 (5.3\%) \\
    & \textbf{T5-Extension} & 140 (9.0\%) & 57 (10.7\%) & 107 (21.7\%) & 68 (10.7\%) & 462 (8.2\%) & 0 (0\%) & 33 (11.8\%) & 50 (1.6\%) & 67 (9.9\%) \\
    & \textbf{integration} & 277 (17.8\%) & 57 (10.7\%) & 163 (33.1\%) & 164 (25.7\%) & 689 (12.3\%) & 15 (12.0\%) & 77 (27.6\%) & 258 (8.4\%) & 125 (18.5\%) \\
    & \textbf{SAGA} & 364 (23.5\%) & 153 (28.8\%) & 191 (38.7\%) & 165 (25.9\%) & 1093 (19.5\%) & 16 (12.8\%) & 96 (34.4\%) & 350 (11.4\%) & 136 (20.1\%) \\
    \hline\hline\hline
  \end{tabular}}
}
\end{table*}
\baselineskip=18pt plus.2pt minus.2pt
\parskip=0pt plus.2pt minus0.2pt

\subsubsection{Experimental Metrics Evaluation}

Table~\ref{rq1a} shows the model performance on the two datasets in terms of the three evaluation metrics. The best result for each metric is marked in bold. As shown in Table~\ref{rq1a}, SAGA substantially outperforms the five baselines on both datasets. Specifically, SAGA achieves an accuracy of 53.1\% and 19.8\% on the two datasets, respectively, which achieves a relative improvement of 42.7\% and 40.4\% over the best baseline model integration. In terms of the BLEU-4 metric, SAGA obtains 75.56 and 39.17 scores on the two datasets, respectively, which is 15.83 and 16.39 points higher than integration. In terms of the ROUGE-L metric, SAGA obtains 85.96 and 65.15 scores on the two datasets, respectively, which is 6.45 and 2.89 points higher than integration.

In addition, we observe that T5 yields poor performance on the two datasets compared with the results in the original paper \cite{26}. Since we use the public source code provided by the authors and follow the same training strategy as in the original paper, we further look into the datasets and draw the following possible reasons: 1) On $\mathbf{CAPS}_A$ and $\mathbf{CAPS}_M$, the average length of the assert statements (after adding the complete package name to the front of each declared class) in the testing set is 21.69 and 22.32 tokens, respectively, while the average length is 17.25 tokens in T5's testing set. Thus, the need for generating longer assert statements may be a reason for decreasing the model performance. Furthermore, because the complete package name list typically consists of fixed patterns, T5 achieves a comparable BLEU-4 score while maintaining low accuracy; 2) Since $\mathbf{CAPS}_A$ and $\mathbf{CAPS}_M$ contain the challenging cases of assert statements with unknown tokens, however, T5 is fine-tuned on the dataset that excludes such challenging cases. Therefore, T5 may be less capable of dealing with open-vocabulary issues; 3) After analyzing the vocabulary of T5 dataset, we discover that all uppercase letters are replaced with lowercase letters. Such modifications would negatively change the tokens with different semantics into the same and reduce the difficulty of this task as well.

\setcounter{figure}{4}
\begin{figure*}[b]
\centering
  \subfigure[]{
    \includegraphics[width=0.48\textwidth]{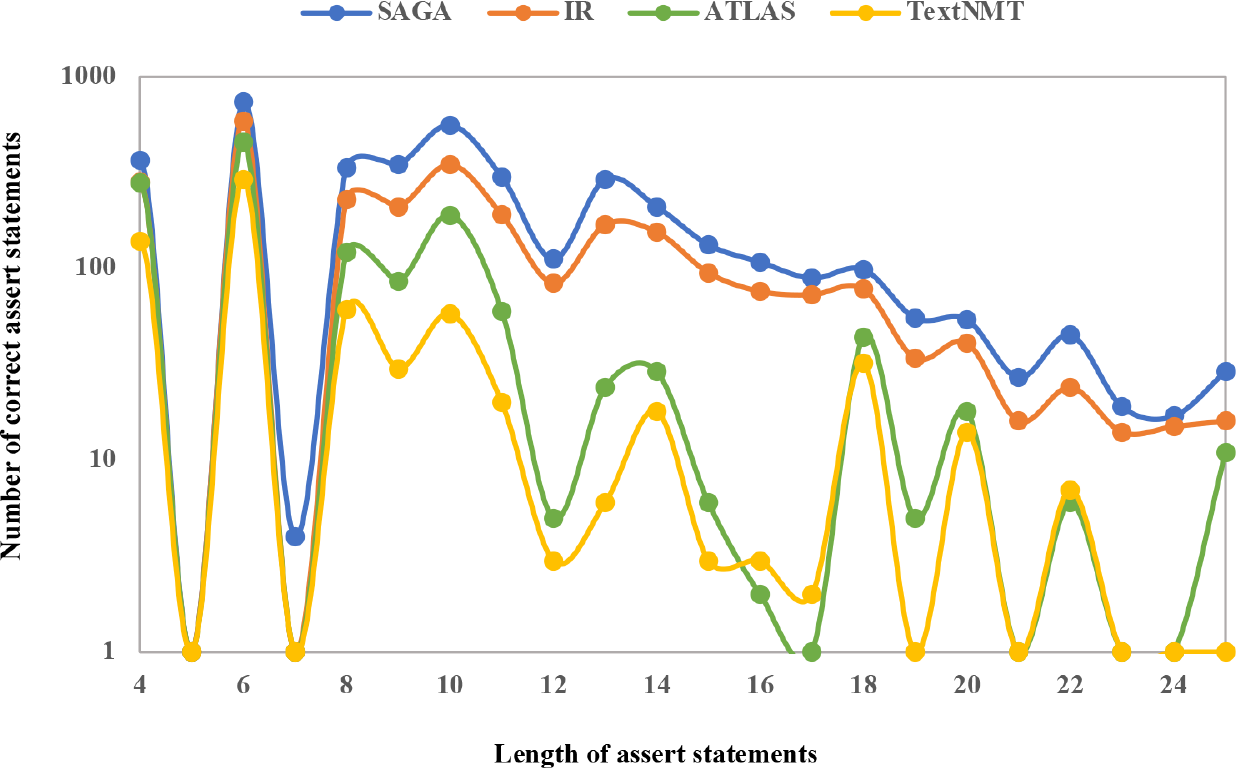}}
  \subfigure[]{
    \includegraphics[width=0.48\textwidth]{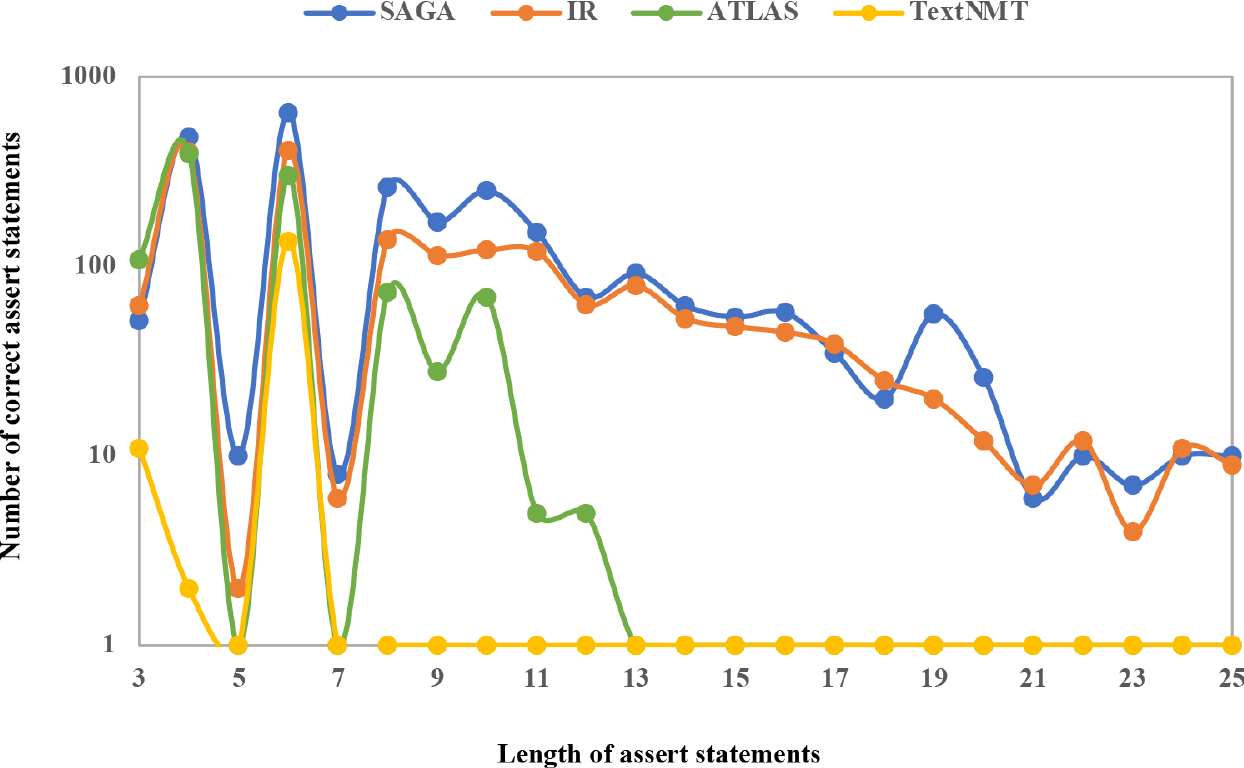}}
  \caption{Length distribution of the correct assert statements. (a) $\mathbf{CAPS}_A$. (b) $\mathbf{CAPS}_M$.}
  \label{length}
\end{figure*}

\subsubsection{Assert Statement Types Evaluation}

In addition, we also analyze the type of assert statement that are correctly generated by each model. Table~\ref{rq1b} presents the accuracy results on assert statements of different types (with the number of exact matches). Note that the last column indicates the results of four other assert statement types (i.e., \texttt{assertNotEquals}, \texttt{assertNotSame}, \texttt{assertThrows} and \texttt{fail}) that are not included in the $\mathbf{CAPS}_A$ dataset. As it can be seen in Table~\ref{rq1b}, SAGA is able to consistently outperform the baselines in all the assert statement types on both datasets. The distribution of each type correctly generated by SAGA is relatively even, which mitigates the possible threat that SAGA is only capable of generating a specific type of assert statement. Watson  {\it et al.} \cite{8} hypothesize that the \texttt{assertThat} statements are more difficult to generate due to the nature of the assert itself. Despite the complexities of \texttt{assertThat} statements, as the developer-written summarization often contains explicit hints about how to understand the intended functionality of focal method, SAGA is able to achieve high prediction accuracy of 50.7\% on the $\mathbf{CAPS}_A$ dataset. As for the challenging dataset $\mathbf{CAPS}_M$, SAGA can still correctly predict 11.4\% of the assert statements in the testing set.

\setcounter{table}{3}
\tabcolsep 9pt
\renewcommand\arraystretch{1.3}
\begin{table*}[t]
\centering
\caption{\label{rq1c} Statistic Results of the Lengths of the Generated Correct Assert Statements}\vspace{-2mm}
{\footnotesize
  \begin{tabular}{lrrcrrc}
    \hline\hline\hline
    \textbf{Model} & \multicolumn{3}{c}{$\mathbf{CAPS}_A$} & \multicolumn{3}{c}{$\mathbf{CAPS}_M$} \\
    \cmidrule(rl){2-4}
    \cmidrule(rl){5-7}
    & \multicolumn{1}{c}{$\mathbf{Mean}_S$} & \multicolumn{1}{c}{$\mathbf{Mean}_L$} & \multicolumn{1}{c}{\textbf{Median}} & \multicolumn{1}{c}{$\mathbf{Mean}_L$} & \multicolumn{1}{c}{$\mathbf{Mean}_L$} & \multicolumn{1}{c}{\textbf{Median}} \\
    \hline
    \textbf{TestNMT} & 6.75 (11.2\%) & 16.64 (3.4\%) & 6 & 5.75 (1.6\%) & 0 (0\%) & 6 \\
    \textbf{ATLAS} & 7.14 (22.3\%) & 17.28 (5.2\%) & 6 & 5.44 (10.7\%) & 0 (0\%) & 4 \\
    \textbf{integration} & 8.56 (40.3\%) & 18.05 (27.2\%) & 10 & 7.32 (17.2\%) & 17.95 (7.3\%) & 8 \\
    \textbf{SAGA} & 8.79 (58.3\%) & 18.18 (38.0\%) & 10 & 7.43 (24.7\%) & 18.01 (9.1\%) & 8 \\
    \hline\hline\hline
  \end{tabular}
}
\end{table*}
\baselineskip=18pt plus.2pt minus.2pt
\parskip=0pt plus.2pt minus0.2pt

\subsubsection{Correct Assert Statement Length Distribution Evaluation}

We further investigate the ability of each model to correctly predict long assert statements by analyzing the length distribution of generated assert statements. Fig.\ref{length} shows the length distribution of correct assert statements generated by each model on the two datasets, where the X-axis represents the length of assert statements (i.e., the number of tokens within each assert statement) and the Y-axis represents the number of correct assert statements for each corresponding scale on the X-axis. We exclude the two models (T5 and T5-Extension) with different assert statement length in this comparison experiment. As shown in Fig.\ref{length}, it is notable that SAGA tends to be superior to all the baselines in generating both short and long assert statements.

\setcounter{table}{4}
\tabcolsep 9pt
\renewcommand\arraystretch{1.3}
\begin{table*}[b]
\centering
\caption{\label{rq2a} Comparison Results of the Two Metrics for RQ2}\vspace{-2mm}
{\footnotesize
  \begin{tabular}{lrrrrrrr}
    \hline\hline\hline
    \multicolumn{2}{c}{\textbf{Model}} & \multicolumn{3}{c}{$\mathbf{CAPS}_A$} & \multicolumn{3}{c}{$\mathbf{CAPS}_M$} \tabularnewline
    \cmidrule(rl){3-5}
    \cmidrule(rl){6-8}
    & & \multicolumn{1}{c}{\textbf{Accuracy}} & \multicolumn{1}{c}{\textbf{BLEU-4}} & \multicolumn{1}{c}{\textbf{ROUGE-L}} & \multicolumn{1}{c}{\textbf{Accuracy}} & \multicolumn{1}{c}{\textbf{BLEU-4}} & \multicolumn{1}{c}{\textbf{ROUGE-L}} \\
    \hline
    \textbf{TestNMT} & w/o S & 9.5\% & 21.74 & 60.95 & 1.1\% & 2.87 & 46.04 \\
    & w/ S & 12.6\% & 25.91 & 61.52 & 4.9\% & 12.13 & 52.12 \\
    \textbf{ATLAS} & w/o S & 18.0\% & 28.70 & 70.01 & 7.6\% & 14.22 & 60.36 \\
    & w/ S & 23.8\% & 39.89 & 74.74 & 9.9\% & 17.94 & 61.90 \\
    \textbf{T5} & w/o S & 9.1\% & 26.44 & 43.94 & 2.1\% & 20.31 & 49.74 \\
    & w/ S & 9.8\% & 27.52 & 44.20 & 3.2\% & 23.24 & 52.45 \\
    \textbf{T5-Extension} & w/o S & 23.8\% & 33.02 & 72.15 & 7.6\% & 21.86 & 57.85 \\
    & w/ S & 24.7\% & 41.45 & 73.36 & 8.1\% & 23.61 & 59.56 \\
    \textbf{integration} & w/o S & 37.2\% & 59.73 & 79.51 & 14.1\% & 22.78 & 62.26 \tabularnewline
    & w/ S & 37.4\% & 60.92 & 80.13 & 14.1\% & 22.94 & 62.56 \\
    \textbf{SAGA} & w/o S & 52.7\% & 75.28 & 85.53 & 19.3\% & 38.10 & 64.69 \\
    & w/ S & 53.1\% & 75.56 & 85.96 & 19.8\% & 39.17 & 65.15 \\
    \hline\hline\hline
  \end{tabular}
}
\end{table*}
\baselineskip=18pt plus.2pt minus.2pt
\parskip=0pt plus.2pt minus0.2pt

Table~\ref{rq1c} presents the average lengths of short (denoted as $\mathbf{Mean}_S$) and long (denoted as $\mathbf{Mean}_L$) assert statements generated by each model together with the corresponding accuracy, and the median values of all correct assert statements. We regard the assert statements with less than 15 tokens as short in this paper. The statistic results shown in Table~\ref{rq1c} validate our observation that SAGA is capable of correctly generating both short and long assert statements on the two datasets. In addition, the IR-based model is able to retrieve long sequences from the training corpus, integration thus achieves a comparable result against SAGA.

\subsubsection{Answer to RQ1}

In summary, the proposed SAGA framework significantly outperforms the baselines in terms of all the experimental metrics. Our observations indicate that SAGA is capable of generating both long assert statements and the challenging cases with higher accuracy against the baselines.

\subsection{Answering RQ2}

To answer this question, we evaluate the effectiveness of developer-written summarization by conducting ablation experiments on each model (i.e., training the corresponding model with summarization or not) separately. For a fair comparison, the training strategy and the hyper-parameter settings are consistent with those described in Section~\ref{design}.

\subsubsection{Ablation Study}
\label{rq2222}

Table~\ref{rq2a} presents the comparison results of the ablation study. Each model comprises two lines of experimental results, in which the first line shows the results of the model that is trained without using the developer-written summarization and the second line shows the results of using such additional information. As shown in Table~\ref{rq2a}, we can observe that providing the summarization as complementary information contributes to improving the performance of all the models.

We also statistically compare the performance of two different treatments (i.e., with and without summarization) in terms of \textbf{Accuracy} for each corresponding model using the McNemar's test \cite{36}, which is a non-parametric statistical test suitable to the paired dichotomous data summarized in a contingency table \cite{37}. To compute the test results for two treatments $T_{1}$ (i.e., with summarization) and $T_{2}$ (i.e., without summarization), we firstly construct a contingency table by counting the number of cases in which (1) both $T_{1}$ and $T_{2}$ generate the correct assert statement, (2) only $T_{1}$ generates the correct assert statement, (3) only $T_{2}$ generates the correct assert statement, and (4) neither $T_{1}$ nor $T_{2}$ generate the correct assert statement. Then, the McNemar's test is applied to the constructed contingency table to check the null hypothesis stating that the difference between two treatments is insignificant. If the reported \textit{p}-value is less than the significant level 0.05, the null hypothesis will be rejected, and it is drawn that the disparity between treatments is significant and not random. The implementation of the McNemar's test is available at the \texttt{mcnemar} function of the \texttt{mlxtend} Python library \cite{38}. To further complement the results of McNemar's test, we use the \texttt{odds\_ratio} Python library$^{\footnotesize\textcircled{\tiny3}}$ to compute the odd ratio (OR) for measuring the effect size. The OR value greater than 1 means the usage of augmented information has a positive relationship with the generation of meaningful assert statements (i.e., more assert statements could be correctly generated with the aid of providing additional summarization).

\let\thefootnote\relax\footnotetext{{}\indent\ $^{\footnotesize\textcircled{\tiny3}}$https://github.com/JiguangPeng/odds\_ratio, Jan. 2023.}

Table 6 reports the results of McNemar's test to determine if there are statistical differences when training models with the two different treatments. The following results are the observations from Table 6:
\begin{itemize}
	\item As for the five DL-based models (TestNMT, ATLAS, T5, T5-Extension and SAGA), $T_{1}$ leads to significantly better results (\textit{p}-value $<$ 0.05) with the values of OR ranging from 1.03 to 1.43. This means that chances of generating a correct assert statement using $T_{1}$ are 3\% to 43\% higher when compared to $T_{2}$.
	\item As for the IR-based model integration, we can see that there is no statistically significant difference between $T_{1}$ and $T_{2}$ (\textit{p}-value is greater than 0.05). Nevertheless, the value of OR (i.e., 1.01) indicates that $T_{1}$ still improves the performance of integration to some extent.
	\item In view of the inconsistent results described above, we give the following possible explanations: 1) Intuitively, the DL-based models are able to directly learn definitive information from the provided summarization to aid the assert statement generation task; 2) As the key technique of integration is the IR-based assertion retrieval, which is based on the Jaccard similarity between the corresponding and given focal-test written in PL, solely providing additional summarization written in NL is difficult to continue to increase the performance improvements during the retrieval process.
\end{itemize}

\setcounter{figure}{5}
\begin{figure*}[t]
  \centering
  \includegraphics[width=\textwidth]{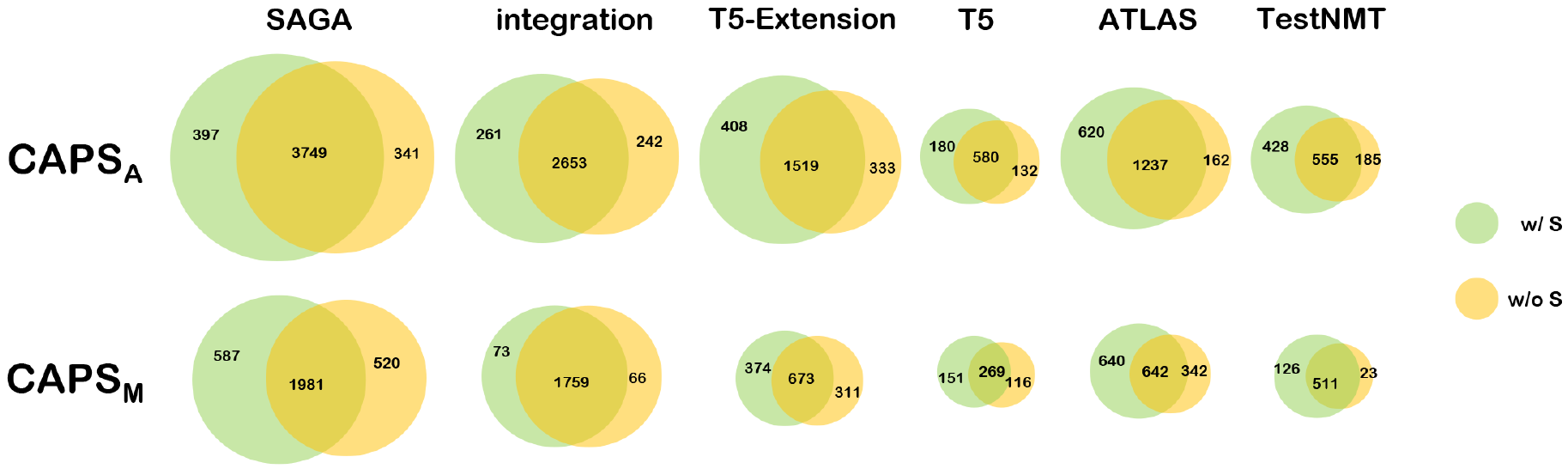}
  \caption{Overlapping of the correct assert statements.}
  \label{fig6}
\end{figure*}

\tabcolsep 10pt
\renewcommand\arraystretch{1.3}
\begin{center}
{\footnotesize{\bf Table 6.} The McNemar's Test (adj. \textit{p}-value and OR) in terms of the Accuracy Metric for RQ2}\\
\vspace{2mm}
\footnotesize{
\begin{tabular*}{\linewidth}{lrcrc}
\hline\hline\hline
    \textbf{Model} & \multicolumn{2}{c}{$\mathbf{CAPS}_A$} & \multicolumn{2}{c}{$\mathbf{CAPS}_M$} \\
    \cmidrule(rl){2-3}
    \cmidrule(rl){4-5}
    & \multicolumn{1}{c}{\textbf{\textit{p}-value}} & \multicolumn{1}{c}{\textbf{OR}} & \multicolumn{1}{c}{\textbf{\textit{p}-value}} & \multicolumn{1}{c}{\textbf{OR}} \\
    \hline
    \textbf{TestNMT} & $<$ 0.05 & 1.38 & $<$ 0.05 & 1.21 \\
    \textbf{ATLAS} & $<$ 0.05 & 1.43 & $<$ 0.05 & 1.36 \\
    \textbf{T5} & $<$ 0.05 & 1.08 & $<$ 0.05 & 1.10 \\
    \textbf{T5-Extension} & $<$ 0.05 & 1.06 & $<$ 0.05 & 1.07 \\
    \textbf{integration} & 0.33 & 1.01 & 0.61 & 1.01 \\
    \textbf{SAGA} & $<$ 0.05 & 1.03 & $<$ 0.05 & 1.04 \\
    \hline\hline\hline
\end{tabular*}
}
\end{center}

\setcounter{figure}{6}
\begin{figure*}[b]
\centering
  \subfigure[]{
    \includegraphics[width=0.48\textwidth]{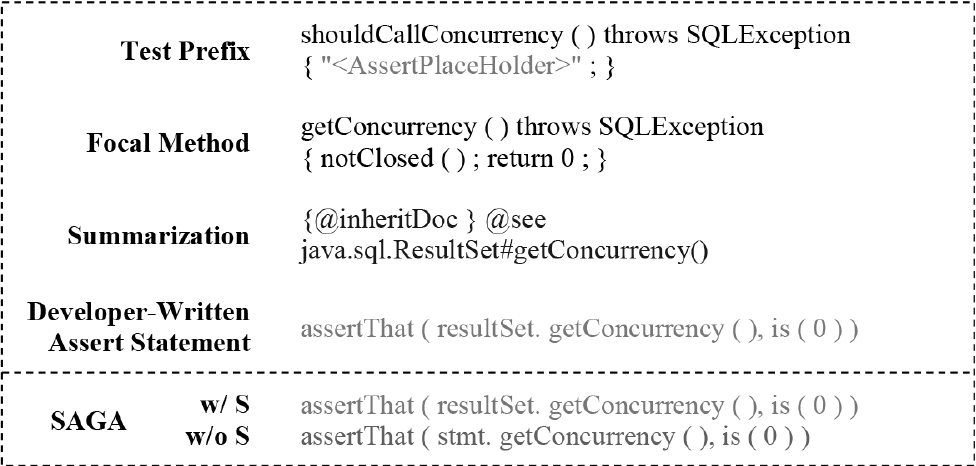}}
  \subfigure[]{
    \includegraphics[width=0.48\textwidth]{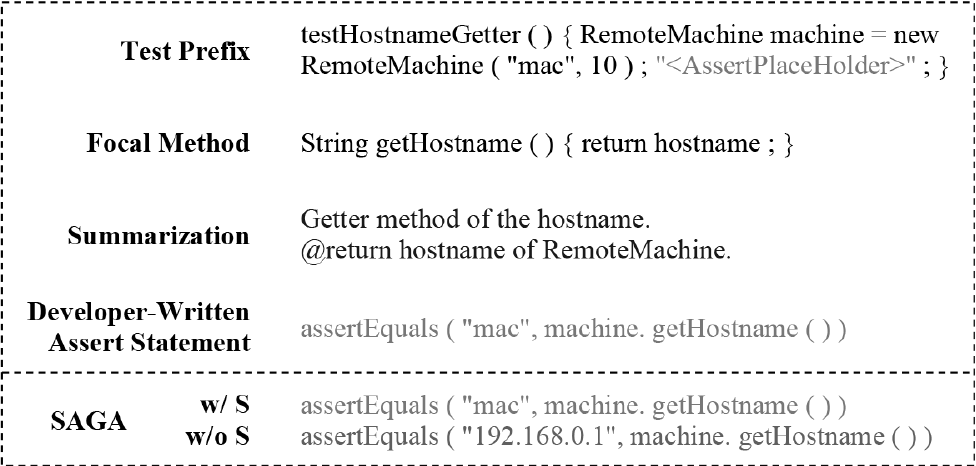}}
  \caption{Examples showing the effect of summarization in SAGA's performance. (a) An example from $\mathbf{CAPS}_A$ dataset. (b) An example from $\mathbf{CAPS}_M$ dataset.}
  \label{motirq2}
\end{figure*}

Additionally, we analyze the uniqueness of correct assert statements generated by each model. Fig.\ref{fig6} shows the overlapping between the correct assert statements generated by using input with summarization or not among each model evaluated on the two datasets. As shown in Fig.\ref{fig6}, we can find that the summarization-guided models (colored with light green) tend to generate more unique correct assert statements that fail to be generated by models without using summarization. For example, 587 correct assert statements are uniquely generated by SAGA on $\mathbf{CAPS}_M$, while 520 correct assert statements are uniquely generated by SAGA without using summarization. By further investigating the incorrect assert statements generated by SAGA, we observe the existence of equivalent cases that are not exactly matched with the developer-written ones but semantically equivalent to the developer's intent. We will discuss these cases in Section~\ref{ae}.

\subsubsection{Case Study}

To better understand the effectiveness of using summarization as complimentary information, we present two cases in Fig.\ref{motirq2} to demonstrate the ability of summarization to guide the generation of assert statements. As a case study, we take the SAGA model as an example to show the difference between assert statements generated with or without summarization. Fig.\ref{motirq2}(a) shows an example of summarization explicitly providing the related token (i.e., \texttt{resultSet}), which is missing from the test prefix and focal method. Fig.8 visualizes the attention weights for the encoder and decoder while generating the expected assert statement. We can observe that SAGA learns the relationship that function \texttt{getConcurrency} belongs to class \texttt{ResultSet} from summarization and thus correctly predicts the token \texttt{resultSet} as the parameter for the assert statement. Nevertheless, SAGA predicts an irrelevant token \texttt{stmt} when the summarization is not used. Fig.\ref{motirq2}(b) depicts an example of summarization used to convey the intended functionality of the focal method. From the content of the summarization, we can clearly understand that the focal method completes the functionality of returning the \texttt{hostname} of \texttt{RemoteMachine}. As shown in Fig.9, SAGA indeed learns the relationship that the hostname of \texttt{machine} is \texttt{"mac"}. Likewise, SAGA fails to predict the token if the summarization is not available. These results reveal the effectiveness of our proposed approach for assert statement generation.

\subsubsection{Answer to RQ2}

To sum up, providing the developer-written summarization can improve the performance of the assert statement generation task. Specifically, the summarization contents may explicitly contain the related tokens directly appearing as parameters in the assert statements or convey the intended program behavior via detailed functionality descriptions of the focal methods to assist in the generation of correct assert statements.

\begin{center}
\includegraphics[width=0.48\textwidth]{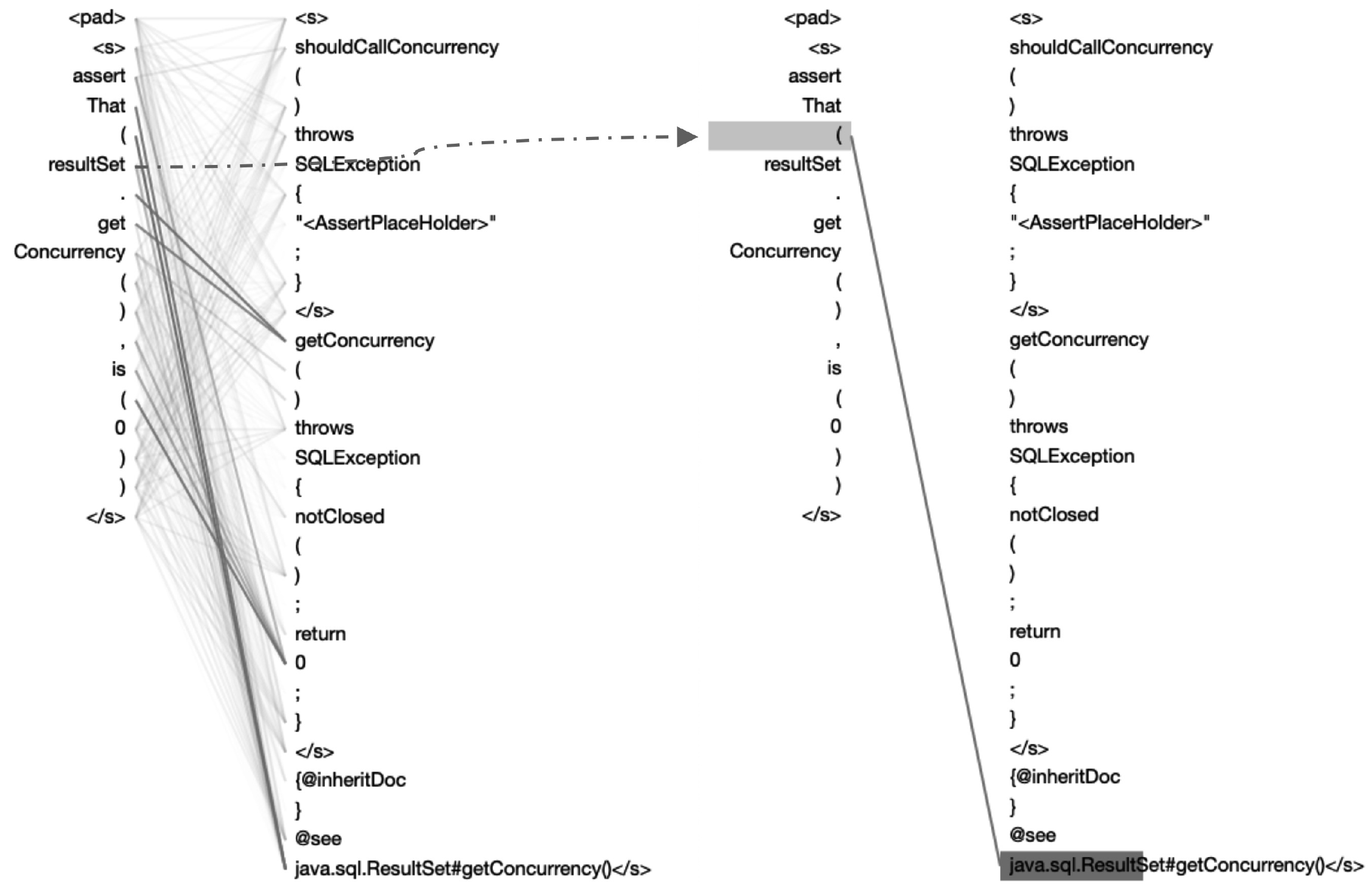}\\
\vspace{2mm}
\parbox[c]{8.3cm}{\footnotesize{Fig.8.~}Visualization of attention weights for example in Fig.7(a).}
\label{case1}
\end{center}

\begin{center}
\includegraphics[width=0.48\textwidth]{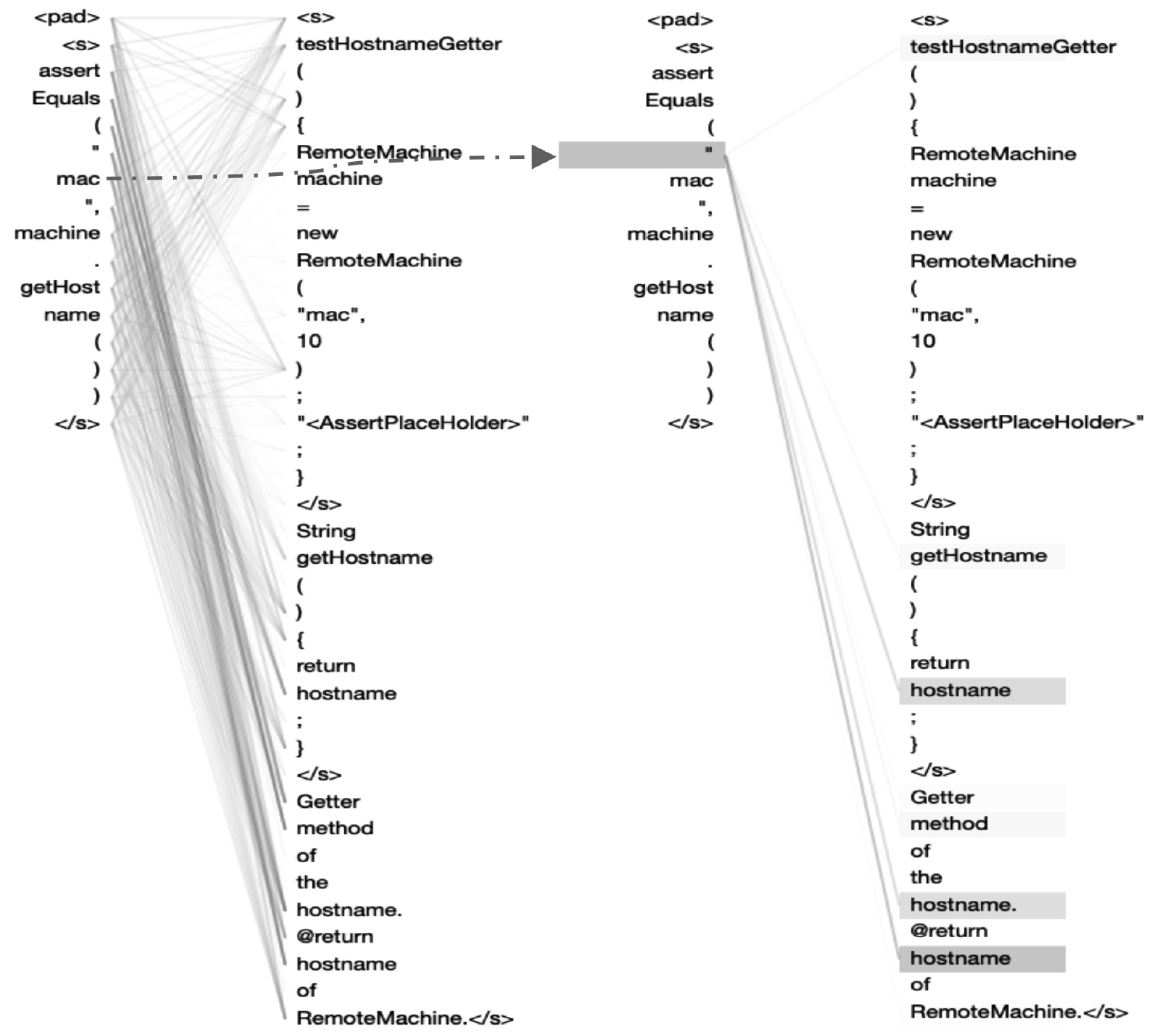}\\
\vspace{2mm}
\parbox[c]{8.3cm}{\footnotesize{Fig.9.~}Visualization of attention weights for example in Fig.7(b).}
\label{case2}
\end{center}

\subsection{Answering RQ3}

To answer this research question, we inspect the assert statements that are not exactly matched with the ground truth. The evaluation is split into two aspects: 1) discussing the semantically equivalent examples in the incorrect assert statements generated by SAGA; 2) calculating the edit distance of incorrect assert statements generated by SAGA.

\setcounter{table}{6}
\tabcolsep 9pt
\renewcommand\arraystretch{1.3}
\begin{table*}[t]
\centering
\caption{\label{rq3a} Comparison Results of Edit Distance Analysis}\vspace{-2mm}
{\footnotesize
  \resizebox{\textwidth}{!}{
  \begin{tabular}{crrrrrrrrrrrr}
    \hline\hline\hline
    \textbf{Edit Dist.} & \multicolumn{6}{c}{$\mathbf{CAPS}_A$} & \multicolumn{6}{c}{$\mathbf{CAPS}_M$} \\
    \cmidrule(rl){2-7}
    \cmidrule(rl){8-13}
    & \multicolumn{1}{c}{\textbf{TestNMT}} & \multicolumn{1}{c}{\textbf{ATLAS}} & \multicolumn{1}{c}{\textbf{T5}} & \multicolumn{1}{c}{\textbf{T5-Extension}} & \multicolumn{1}{c}{\textbf{integration}} & \multicolumn{1}{c}{\textbf{SAGA}} & \multicolumn{1}{c}{\textbf{TestNMT}} & \multicolumn{1}{c}{\textbf{ATLAS}} & \multicolumn{1}{c}{\textbf{T5}} & \multicolumn{1}{c}{\textbf{T5-Extension}} & \multicolumn{1}{c}{\textbf{integration}} & \multicolumn{1}{c}{\textbf{SAGA}} \\
    \hline
    \textbf{1} & 732 (10.4\%) & 1085 (17.0\%) & 712 (10.1\%) & 923 (15.5\%) & 1352 (27.6\%) & 1123 (30.7\%) & 261 (2.0\%) & 1177 (9.8\%) & 157 (1.3\%) & 640 (5.3\%) & 1060 (9.7\%) & 1089 (10.5\%) \\
    \textbf{2} & 784 (11.1\%) & 583 (9.1\%) & 183 (2.6\%) & 451 (7.6\%) & 529 (10.8\%) & 490 (13.4\%) & 704 (5.5\%) & 751 (6.3\%) & 254 (2.0\%) & 451 (3.8\%) & 837 (7.7\%) & 893 (8.6\%) \\
    \textbf{3} & 691 (9.8\%) & 573 (9.0\%) & 59 (0.8\%) & 506 (8.5\%) & 444 (9.1\%) & 355 (9.7\%) & 555 (4.3\%) & 888 (7.4\%) & 204 (1.6\%) & 629 (5.2\%) & 775 (7.1\%) & 794 (7.6\%) \\
    \hline\hline\hline
  \end{tabular}}
}
\end{table*}
\baselineskip=18pt plus.2pt minus.2pt
\parskip=0pt plus.2pt minus0.2pt

\subsubsection{Equivalence Evaluation}
\label{ae}

In this section, we manually analyze the incorrect assert statements generated by SAGA and present qualitative discussion. As shown in Fig.10, the list of equivalent examples showcase some of the SAGA-generated assert statements that do not exactly match with the ground truth, but they are semantically equivalent to the developer-written ones. For example, the developer checks that \texttt{t.getCount() == 999} is true, while SAGA suggests an equivalent check with \texttt{assertEquals(999, t.getCount())}. Similarly, SAGA suggests to assert a null string by checking whether the length of string is equal to 0, while the developer uses the \texttt{""} string directly. In another instance, SAGA suggests to use the \texttt{assertEquals} statement to judge the equivalence of two objects rather than the \texttt{assertThat} statement. The last two instances show that SAGA is able to successfully predict the full assert statements except the given message strings (one is different and the other is missing). Given that the message strings do not provide crucial logic checks in the test cases, these instances are still valuable for the developers.

\begin{center}
\includegraphics[width=0.48\textwidth]{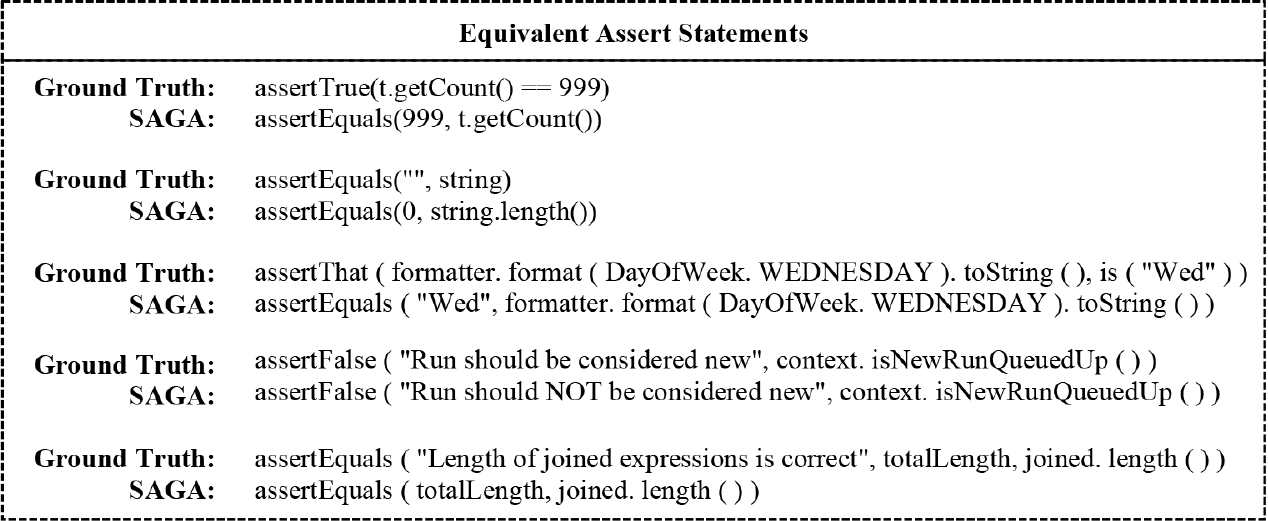}\\
\vspace{2mm}
\parbox[c]{8.3cm}{\footnotesize{Fig.10.~}Examples of equivalent cases generated by SAGA.}
\label{rq3ac}
\end{center}

The existence of equivalence results highlight the need for additional reasonable metrics beyond simple accuracy. In particular, metrics that can recognize cases where the generated assert statement is different yet equivalent to the one written by developers, as well as the non-equivalent ones that can also successfully pass the given unit test and cover the focal method.

\subsubsection{Edit Distance Evaluation}

\let\thefootnote\relax\footnotetext{{}\indent\ $^{\footnotesize\textcircled{\tiny4}}$https://github.com/gfairchild/pyxDamerauLevenshtein, May. 2022.}

This evaluation computes the absolute token-based edit distance between the incorrect assert statements and the manually written ground truth (i.e., the minimum number of operations required to transform incorrect assert statements into correct assert statements). The edit distance metric gives evidence to how useful incorrect assert statements are to developers. Intuitively, the easier it is to transform an incorrect assert statement into a correct assert statement, the more useful that assert statement would be for developers. This evaluation is conducted by using the assert statements generated for RQ1. We find the token-based edit distance by computing the Levenshtein distance between the model-generated and developer-written assert statement using the \texttt{pyxDamerauLevenshtein} library$^{\footnotesize\textcircled{\tiny4}}$. As is shown in Table~\ref{rq3a}, the statistic results reveal that SAGA performs the best in edit distance, with the five baselines trailing behind. Specifically, the number of assert statements that SAGA cannot generate correctly on $\mathbf{CAPS}_A$ and $\mathbf{CAPS}_M$ is 3653 and 10403, respectively. When the edit distance is 1, there are 1123 (i.e., 30.7\%) and 1089 (10.5\%) incorrect assert statements can be converted into correct assert statements on the two datasets, while 1968 (53.8\%) and 2776 (26.7\%) assert statements have a edit distance that no more than 3 tokens from the correct assert statements. In summary, there is a considerable amount of incorrect assert statements generated by SAGA are similar to the developer-written ground truth. Many incorrect results can be turn into perfect predictions by modifying only one token (e.g., related constant or the assert statement type). Thus, these incorrect assert statements can also be useful to aid the developers.

\section{Threats to Validity}
\label{thr}

In this section, we illustrate the main threats to the validity of our approach, which are listed as follows:
\begin{itemize}
	\item \textbf{External threat}: The quality of the datasets is the principal threat to external validity in this paper. We create the CAPS dataset by modifying two existing datasets \cite{8,11}, which are all collected from open-source GitHub repositories. During the construction process, we use some heuristic rules to identify the focal methods for a given test method. Although we did a rigorous data processing, there may be still some noise. In our future research, we will adopt more optimal ways of establishing test-to-code traceability links \cite{39,40} for identifying focal methods more precise. SAGA is also limited by its dependency on the usage of merely developer-written test cases for model training. In general, manually written test cases usually have different characteristics against those generated by automated test case generation tools \cite{41}. As a future direction, SAGA could be trained on an extended dataset consisting of test cases automatically generated by tools, which more closely fits the distribution of tool-generated testing set.
	\item \textbf{Internal threat}: It is widely known that DL-based models are sensitive to hyper-parameters. Thus using a sub-optimal hyper-parameter can pose a internal threat to the validity of SAGA. Due to the limitation of computational resources, we cannot conduct a thorough exploration of optimal hyper-parameters in this paper. Since Raffel  {\it et al.} \cite{10} have explored effective settings of hyper-parameters through extensive experiments in previous work,  we use the exact same hyper-parameters described by their paper. We acknowledge that there might be room for further improvement by additional tuning. It is noted that the large-scale language models (e.g., Codex \cite{42}) trained for code completion are not included as baselines in this paper, since SAGA has a much smaller model size of 60M parameters than theirs of 12B ($\sim$200$\times$). We will further conduct an empirical study on the effectiveness of these general-purpose models on the task of assert statement generation as future work.
	\item \textbf{Construct threat}: In this paper, the experimental metrics used to evaluate model performance are referred to as the construct threat. We adopt three metrics that have been used in previous studies \cite{7,8,29}. Although these metrics do not represent human judgment, they can be used to quickly and quantitatively evaluate the model performance. In the future, we will conduct more human evaluations of the models.
\end{itemize}

\section{Conclusion and Future Work}
\label{con}

In this study, we propose a novel DL-based approach SAGA for assert statement generation. To accurately reflect the developer's intent, we make the first attempt to leverage the summarization of the focal method as complementary information. We then take advantage of a state-of-the-art encoder-decoder language model to automatically generate meaningful assert statements. Empirical results demonstrate that the developer-written summarization can provide definitive information for improving the performance of assert statement generation, thus outperforming the state-of-the-art approaches in terms of all the experimental metrics.

In the future, we plan to use static analysis tools to collect additional contextual information (e.g., global context at project-level) pertinent to the given focal method, aiming to assist SAGA in generating more precise assert statements by augmenting the focal context input. Furthermore, semi-supervised pre-training on projects where SAGA will be used to infer assert statements could help our proposed model to familiarize with project-related knowledge. As discussed earlier, we foresee that such DL-based approaches could be used to support developers in writing unit test cases more efficiently in practice. In this scenario, we consider integrating SAGA as an IDE plugin, which can be regarded as a code completion tool by automatically suggesting assert statements while manually writing unit test cases.

\label{last-page}
\end{multicols}
\label{last-page}

\begin{thebibliography}{99}
\footnotesize
\itemsep=-3pt plus.2pt minus.2pt
\baselineskip=14pt plus.2pt minus.2pt
\bibitem{1} Garousi V, Zhi J. A survey of software testing practices in Canada. {\it J. Syst. Softw.}, 2013, 86(5): 1354--1376. DOI: 10.1016/j.jss.2012.12.051.

\bibitem{2} Pacheco C, Ernst M D. Randoop: Feedback-directed random testing for Java. In {\it Proc. of the Companion to the 22nd Annual {ACM} {SIGPLAN} Conf. on Object-Oriented Programming, Systems, Languages, and Applications}, Oct. 2007, pp.815--816. DOI: 10.1145/1297846.1297902.

\bibitem{3} Pacheco C, Lahiri S K, Ernst M D, Ball T. Feedback-directed random test generation. In {\it Proc. of the 29th Int. Conf. on Software Engineering}, May. 2007, pp.75--84. DOI: 10.1109/ICSE.2007.37.

\bibitem{4} Fraser G, Arcuri A. Evosuite: Automatic test suite generation for object-oriented software. In {\it Proc. of the 19th {ACM} {SIGSOFT} Symp. on the Foundations of Software Engineering and 13th European Software Engineering Conf.}, Sept. 2011, pp.416--419. DOI: 10.1145/2025113.2025179.

\bibitem{5} Shamshiri S. Automated unit test generation for evolving software. In {\it Proc. of the 2015 10th Joint Meeting on Foundations of Software Engineering}, Aug. 2015, pp.1038--1041. DOI: 10.1145/2786805.2803196.

\bibitem{6} Almasi M M, Hemmati H, Fraser G, Arcuri A, Benefelds J. An industrial evaluation of unit test generation: Finding real faults in a financial application. In {\it Proc. of 39th {IEEE/ACM} Int. Conf. on Software Engineering: Software Engineering in Practice Track}, May. 2017, pp.263--272. DOI: 10.1109/ICSE-SEIP.2017.27.

\bibitem{7} White R, Krinke J. TestNMT: Function-to-test neural machine translation. In {\it Proc. of the 4th {ACM} {SIGSOFT} Int. Workshop on {NLP} for Software Engineering}, Nov. 2018, pp.30--33. DOI: 10.1145/3283812.3283823.

\bibitem{8} Watson C, Tufano M, Moran K, Bavota G, Poshyvanyk D. On learning meaningful assert statements for unit test cases. In {\it Proc. of the 42nd Int. Conf. on Software Engineering}, Jun. 2020, pp.1398--1409. DOI: 10.1145/3377811.3380429.

\bibitem{9} Wang Y, Wang W, Joty S R, Hoi S C H. CodeT5: Identifier-aware unified pre-trained encoder-decoder models for code understanding and generation. In {\it Proc. of the 2021 Conf. on Empirical Methods in Natural Language Processing}, Nov. 2021, pp.8696--8708. DOI: 10.18653/v1/2021.emnlp-main.685.

\bibitem{10} Raffel C, Shazeer N, Roberts A, Lee K, Narang S, Matena M, Zhou Y, Li W, Liu P J. Exploring the limits of transfer learning with a unified text-to-text transformer. {\it J. Mach. Learn. Res.}, 2020, 21: 140:1--140:67.

\bibitem{11} Tufano M, Deng S K, Sundaresan N, Svyatkovskiy A. METHODS2TEST: A dataset of focal methods mapped to test cases. In {\it Proc. of the 19th {IEEE/ACM} Int. Conf. on Mining Software Repositories}, May. 2022, pp.299--303. DOI: 10.1145/3524842.3528009.

\bibitem{12} Padhye R, Lemieux C, Sen K. JQF: Coverage-guided property-based testing in Java. In {\it Proc. of the 28th {ACM} {SIGSOFT} Int. Symp. on Software Testing and Analysis}, Jul. 2019, pp.398--401. DOI: 10.1145/3293882.3339002.

\bibitem{13} Gopinath R, Kampmann A, Havrikov N, Soremekun E O, Zeller A. Abstracting failure-inducing inputs. In {\it Proc. of the 29th {ACM} {SIGSOFT} Int. Symp. on Software Testing and Analysis}, Jul. 2020, pp.237--248. DOI: 10.1145/3395363.3397349.

\bibitem{14} Li X, Li W, Zhang Y, Zhang L. DeepFL: Integrating multiple fault diagnosis dimensions for deep fault localization. In {\it Proc. of the 28th {ACM} {SIGSOFT} Int. Symp. on Software Testing and Analysis}, Jul. 2019, pp.169--180. DOI: 10.1145/3293882.3330574.

\bibitem{15} Wang S, Liu T, Tan L. Automatically learning semantic features for defect prediction. In {\it Proc. of the 38th Int. Conf. on Software Engineering}, May. 2016, pp.297--308. DOI: 10.1145/2884781.2884804.

\bibitem{16} Zhang Y, Jin D, Xing Y, Gong Y. Automated defect identification via path analysis-based features with transfer learning. {\it J. Syst. Softw.}, 2020, 166: 110585. DOI: 10.1016/j.jss.2020.110585.

\bibitem{17} Zhao Y, Wang Y, Zhang Y, Zhang D, Gong Y, Jin D. ST-TLF: Cross-version defect prediction framework based transfer learning. {\it Inf. Softw. Technol.}, 2022, 149: 106939. DOI: 10.1016/j.infsof.2022.106939.

\bibitem{18} Xing Y, Qian X, Guan Y, Yang B, Zhang Y. Cross-project defect prediction based on G-LSTM model. {\it Pattern Recognit. Lett.}, 2022, 160: 50--57. DOI: 10.1016/j.patrec.2022.04.039.

\bibitem{19} Luo S, Xu H, Bi Y, Wang X, Zhou Y. Boosting symbolic execution via constraint solving time prediction (experience paper). In {\it Proc. of the 30th {ACM} {SIGSOFT} Int. Symp. on Software Testing and Analysis}, Jul. 2021, pp.336--347. DOI: 10.1145/3460319.3464813.

\bibitem{20} Pan C, Pradel M. Continuous test suite failure prediction. In {\it Proc. of the 30th {ACM} {SIGSOFT} Int. Symp. on Software Testing and Analysis}, Jul. 2021, pp.553--565. DOI: 10.1145/3460319.3464840.

\bibitem{21} Chen J, Bai Y, Hao D, Xiong Y, Zhang H, Xie B. Learning to prioritize test programs for compiler testing. In {\it Proc. of the 39th Int. Conf. on Software Engineering}, May. 2017, pp.700--711. DOI: 10.1109/ICSE.2017.70.

\bibitem{22} Spieker H, Gotlieb A, Marijan D, Mossige M. Reinforcement learning for automatic test case prioritization and selection in continuous integration. In {\it Proc. of the 26th {ACM} {SIGSOFT} Int. Symp. on Software Testing and Analysis}, Jul. 2017, pp.12--22. DOI: 10.1145/3092703.3092709.

\bibitem{23} Lutellier T, Pham H V, Pang L, Li Y, Wei M, Tan L. CoCoNut: Combining context-aware neural translation models using ensemble for program repair. In {\it Proc. of the 29th {ACM} {SIGSOFT} Int. Symp. on Software Testing and Analysis}, Jul. 2020, pp.101--114. DOI: 10.1145/3395363.3397369.

\bibitem{24} Zhu Q, Sun Z, Xiao Y, Zhang W, Yuan K, Xiong Y, Zhang L. A syntax-guided edit decoder for neural program repair. In {\it Proc. of the 29th {ACM} Joint European Software Engineering Conf. and Symp. on the Foundations of Software Engineering}, Aug. 2021, pp.341--353. DOI: 10.1145/3468264.3468544.

\bibitem{25} Chen Z, Kommrusch S, Tufano M, Pouchet L, Poshyvanyk D, Monperrus M. SequenceR: Sequence-to-sequence learning for end-to-end program repair. {\it {IEEE} Trans. Software Eng.}, 2021, 47(9): 1943--1959. DOI: 10.1109/TSE.2019.2940179.

\bibitem{26} Mastropaolo A, Scalabrino S, Cooper N, Nader-Palacio D, Poshyvanyk D, Oliveto R, Bavota G. Studying the usage of text-to-text transfer transformer to support code-related tasks. In {\it Proc. of the 43rd Int. Conf. on Software Engineering}, May. 2021, pp.336--347. DOI: 10.1109/ICSE43902.2021.00041.

\bibitem{27} Mastropaolo A, Cooper N, Nader-Palacio D, Scalabrino S, Poshyvanyk D, Oliveto R, Bavota G. Using transfer learning for code-related tasks. {\it {IEEE} Trans. Software Eng.}, 2023, 49(4): 1580--1598. DOI: 10.1109/TSE.2022.3183297.

\bibitem{28} Dinella E, Ryan G, Mytkowicz T, Lahiri S K. TOGA: A neural method for test oracle generation. In {\it Proc. of the 44th Int. Conf. on Software Engineering}, May. 2022, pp.2130--2141. DOI: 10.1145/3510003.3510141.

\bibitem{29} Yu H, Lou Y, Sun K, Ran D, Xie T, Hao D, Li Y, Li G, Wang Q. Automated assertion generation via information retrieval and its integration with deep learning. In {\it Proc. of the 44th Int. Conf. on Software Engineering}, May. 2022, pp.163--174. DOI: 10.1145/3510003.3510149.

\bibitem{30} Kudo T, Richardson J. SentencePiece: A simple and language independent subword tokenizer and detokenizer for neural text processing. In {\it Proc. of the 2018 Conf. on Empirical Methods in Natural Language Processing}, Oct. 2018, pp.66--71. DOI: 10.18653/v1/d18-2012.

\bibitem{31} Gu J, Lu Z, Li H, Li V O K. Incorporating copying mechanism in sequence-to-sequence learning. In {\it Proc. of the 54th Annual Meeting of the Association for Computational Linguistics}, Aug. 2016. DOI: 10.18653/v1/p16-1154.

\bibitem{32} Pawlak R, Monperrus M, Petitprez N, Noguera C, Seinturier L. SPOON: A library for implementing analyses and transformations of Java source code. {\it Softw. Pract. Exp.}, 2016, 46(9): 1155--1179. DOI: 10.1002/spe.2346.

\bibitem{33} Klein G, Kim Y, Deng Y, Senellart J, Rush A M. OpenNMT: Open-source toolkit for neural machine translation. In {\it Proc. of the 55th Annual Meeting of the Association for Computational Linguistics}, Jul. 2017, pp.67--72. DOI: 10.18653/v1/P17-4012.

\bibitem{34} Papineni K, Roukos S, Ward T, Zhu W. Bleu: A method for automatic evaluation of machine translation. In {\it Proc. of the 40th Annual Meeting of the Association for Computational Linguistics}, Jul. 2002, pp.311--318. DOI: 10.3115/1073083.1073135.

\bibitem{35} Lin C Y. Rouge: A package for automatic evaluation of summaries. In {\it Text Summarization Branches Out}, 2004, pp.74--81.

\bibitem{36} McNemar Q. Note on the sampling error of the difference between correlated proportions or percentages. {\it Psychometrika}, 1947, 12(2): 153--157.

\bibitem{37} Mohammadi M, Atashin A A, Hofman W, Tan Y. Comparison of ontology alignment systems across single matching task via the McNemar's test. {\it {ACM} Trans. Knowl. Discov. Data}, 2018, 12(4): 51:1--51:18. DOI: 10.1145/3193573.

\bibitem{38} Raschka S. MLxtend: Providing machine learning and data science utilities and extensions to Python’s scientific computing stack. {\it J. Open Source Softw.}, 2018, 3(24): 638. DOI: 10.21105/joss.00638.

\bibitem{39} Ghafari M, Ghezzi C, Rubinov K. Automatically identifying focal methods under test in unit test cases. In {\it Proc. of the 15th {IEEE} Int. Working Conf. on Source Code Analysis and Manipulation}, Sept. 2015, pp.61--70. DOI: 10.1109/SCAM.2015.7335402.

\bibitem{40} White R, Krinke J, Tan R. Establishing multilevel test-to-code traceability links. In {\it Proc. of the 42nd Int. Conf. on Software Engineering}, Jun. 2020, pp.861--872. DOI: 10.1145/3377811.3380921.

\bibitem{41} Panichella A, Panichella S, Fraser G, Sawant A A, Hellendoorn V J. Revisiting test smells in automatically generated tests: Limitations, pitfalls, and opportunities. In {\it Proc. of the 36th {IEEE} Int. Conf. on Software Maintenance and Evolution}, Sept. 2020, pp.523--533. DOI: 10.1109/ICSME46990.2020.00056.

\bibitem{42} Chen M, Tworek J, Jun H, Yuan Q, Oliveira Pinto H P, Kaplan J, Edwards H, Burda Y, Joseph N, Brockman G, Ray A, Puri R, Krueger G, Petrov M, Khlaaf H, Sastry G, Mishkin P, Chan B, Gray S, Ryder N, Pavlov M, Power A, Kaiser L, Bavarian M, Winter C, Tillet P, Such F P, Cummings D, Plappert M, Chantzis F, Barnes E, Herbert-Voss A, Guss W H, Nichol A, Paino A, Tezak N, Tang J, Babuschkin I, Balaji S, Jain S, Saunders W, Hesse C, Carr A N, Leike J, Achiam J, Misra V, Morikawa E, Radford A, Knight M, Brundage M, Murati M, Mayer K, Welinder P, McGrew B, Amodei D, McCandlish S, Sutskever I, Zaremba W. Evaluating large language models trained on code. arXiv:2107.03374, 2021. https://arxiv.org/abs/2107.03374, Feb. 2023.

\end{thebibliography}
\end{document}